\documentstyle[aps,eqsecnum,preprint,prd,epsfig,epsf]{revtex}
\begin{document}

\preprint{\vbox{\hbox{ MZ-TH/00-29\hfill}
                \hbox{DSF-2000/23\hfill}}} \vskip 1truecm

\vspace{2cm}

\title{The semileptonic decays of the $B_c$ meson}
\author{{\bf M.\ A.\ Ivanov$^a$, J.\ G.\ K\"{o}rner$^b$, P.\ Santorelli$^c$}}
\address{
\baselineskip=15pt
\hspace{0.1truecm}\\
$^a$Bogoliubov Laboratory of Theoretical Physics,\\
Joint Institute for Nuclear Research, 141980 Dubna, Russia\\
\hspace{.1truecm}\\
$^b$Johannes Gutenberg-Universit\"{a}t, Institut f\"ur Physik,\\
Staudinger Weg 7, D-55099, Mainz, Germany\\
\hspace{.1truecm}\\
$^c$ Dipartimento di Scienze Fisiche,
Universit\`a di Napoli ``Federico II"\\
and INFN Sezione di Napoli,
Via Cintia, I-80126 Napoli, Italy}
\maketitle

\begin{abstract}
We study the semileptonic transitions $B_c\to \eta_c$, $J/\psi$,
$D$, $D^\ast$, $B$, $B^\ast$, $B_s$, $B_s^\ast$ in the framework
of a relativistic constituent quark model. We use  experimental
data on leptonic  $J/\psi$ decay, lattice and QCD sum rule results
on leptonic $B_c$ decay, and experimental data on radiative
$\eta_c$ transitions to adjust the quark model parameters.
We compute all form factors of the above semileptonic $B_c$-transitions
and give predictions for various semileptonic $B_c$ decay modes including
their $\tau$-modes when they are kinematically accessible.
The implications of heavy quark symmetry for the  semileptonic decays
are discussed and are shown to be manifest in our explicit relativistic
quark model calculation. A comparison of our results  with the results
of other calculations  is performed.
\end{abstract}

\addtolength{\jot}{10pt}
\tighten

\date{\today}


\newpage

\section{Introduction}

Recently, the observation of the bottom-charm $B_c$ meson at
Fermilab Tevatron has been reported by the CDF Collaboration
\cite{CDF}. The $B_c$ mesons were found in the analysis of their
semileptonic decays, $B_c^{\pm}\to J/\psi l^{\pm} X$. Values for
the mass and the lifetime of the $B_c$ meson were given as
$M(B_c)=6.40\pm 0.39\pm 0.13$ GeV and
 $\tau(B_c)=0.46^{+0.18}_{-0.16}({\rm stat})\pm 0.03({\rm syst})$ ps,
respectively. The branching fraction for $B_c\to J/\psi\ l\ \nu$
relative to that for $B_c\to J/\psi K$ was found to be

\[
\frac{\sigma (B_{c})\times {\rm Br}(B_{c}\rightarrow J/\psi l\nu
)}{\sigma
(B)\times {\rm Br}(B_{c}\rightarrow J/\psi K)}=0.132_{-0.037}^{+0.041}({\rm %
stat})\pm 0.031({\rm syst})_{-0.020}^{+0.032}\,.
\]

The study of the $B_c$ meson is of great interest due to some of
its outstanding features. It is the lowest bound state of two
heavy quarks (charm and bottom) with open (explicit) flavor that
can be compared with
 the charmonium ($c\bar c$-bound state) and the bottomium
($b\bar b$-bound state) which have hidden (implicit) flavor. The
states with hidden flavor  decay strongly and
electromagnetically whereas the $B_c$-meson decays weakly since it
is below the $B\bar D$-threshold. Naively it might appear that the weak
decays of the $B_c$-meson are similar to those of the $B$ and $D$
mesons. However, the situation is quite  different.
%
%
%
  The new spin-flavor symmetry arises for the systems containing one
  heavy quark when the mass of the heavy quark goes to infinity \cite{IW}.
  It gives some relations between the form factors of the physical processes.
  The deviations from heavy quark symmetry are large for the $D$ meson
  and negligibly small for the $B$ meson.
On the contrary, in the case of the  $B_c$ meson a
consistent heavy quark effective theory (for both constituent
quarks) cannot include the heavy flavor symmetry \cite{static}.
However, the residual heavy quark spin symmetry can be used to
reduce the number of independent semileptonic form factors at
least near the zero recoil point \cite{Jenkins}.

In the naive spectator model, one would expect that
$\Gamma(B_c)\approx \Gamma(B)+\Gamma(D)$ which gives
$\tau(B_c)\approx 0.3$ ps, i.e. 1.5 times less than the central
CDF value.
%
%
The dominance  of the $c \rightarrow s$ transition will have to
be investigated in future analysis when more data becomes available.
%
Thus a reliable evaluation of the long distance contributions is very
important for studying the weak $B_c$ decay properties.

The theoretical status of the $B_c$-meson was reviewed in
\cite{Gersh}. In this paper we focus  on its
exclusive leptonic and semileptonic decays which are  sensitive to
the description of long distance effects and are free of further
assumptions, as  for example, factorization of amplitudes in
non-leptonic processes. Our results on the semileptonic transition
form factors can of course be used for a calculation of the
nonleptonic decays of the $B_c$-meson using the factorization approach.

The exclusive semileptonic and nonleptonic (assuming
factorization) decays of the $B_c$-meson were calculated
before in a potential model
approach \cite{CC}. The binding energy and the wave function of
the $B_c$-meson were computed  by using a  flavor-independent
potential with the parameters fixed by the $c\bar c$ and $b \bar b$
spectra and decays. The same processes were also studied in the
framework of the Bethe-Salpeter equation in \cite{AMV}, and, in the
relativistic constituent quark model formulated on the light-front
in \cite{AKNT}. Three-point sum rules of QCD and NRQCD were analyzed
in \cite{KLO,KKL} to obtain the form factors of the semileptonic  decays
of $B^+_c\to J/\psi(\eta_c)l^+\nu$ and $B^+_c\to B_s(B_s^\ast)l^+\nu$.

As shown by the authors of \cite{Jenkins}, the form factors
parameterizing the $B_c$ semileptonic matrix elements can be
related to a smaller set of form factors  if the decoupling of the spin
of the heavy quarks in $B_c$ and in the mesons produced in the
semileptonic decays is exploited. The reduced form factors can be
evaluated as overlap integral of the meson wave-functions
obtained, for example, using a relativistic potential model. This
was performed in \cite{CF1}, where the $B_c$ semileptonic form
factors were computed and predictions for semileptonic and
non-leptonic decay modes were  given.

In this paper we employ the {\it Relativistic Constituent Quark
Model} (RCQM) \cite{RCQM} for the description of $B_c$
semileptonic meson decays.
The RCQM is based on an effective Lagrangian describing the
coupling of hadrons $H$ to their constituent quarks the coupling
strength of which is determined by the compositeness condition $Z_H=0$
\cite{SWH} where $Z_H$ is the wave function renormalization
constant of the hadron $H$.
%
%
  $Z_H^{1/2}$ is the matrix element between a physical particle
  state and the corresponding bare state. The compositeness condition
  $Z_H=0$ enables us to represent a bound state by introducing
  a quasiparticle interacting with its constituents
  so that the renormalization factor is equal to zero. This does not
  mean that we can solve the QCD bound state equations but
  we are able to show that the condition $Z_H=0$ provides an effective
  and self-consistent way to describe the coupling of the particle to its
  constituents.
One starts with an effective Lagrangian written down in
terms of quark and hadron variables. Then, by using  Feynman
rules, the S-matrix elements describing hadronic interactions are
given in terms of a set of quark diagrams. In particular, the
compositeness condition enables one to avoid a double counting of
hadronic degrees of freedom. This approach is self-consistent and
all calculations of physical observables are straightforward.
There is a small set of  model parameters: the values of the constituent quark
masses and the scale parameters that define the size of the distribution of
the constituent quarks inside a given hadron. This distribution
can be related to the relevant Bethe-Salpeter amplitudes.

The shapes of the  vertex functions and the  quark  propagators can in
principle  be found from an analysis of the Bethe-Salpeter and
Dyson-Schwinger equations, respectively, as done e.g. in \cite{DSE}.
The Dyson-Schwinger equation  has been employed to entail
a unified and uniformly accurate description of light- and
heavy-meson observables \cite{DSEH}.
In this paper we, however, choose a more phenomenological approach
were the vertex function is modelled by a Gaussian form, the size
parameter of which is determined by a fit to the leptonic and radiative
decays of the lowest lying charm and bottom mesons. For the quark propagators
we use the local representation.

The leptonic and semileptonic decays of the lower-lying
pseudoscalar mesons ($\pi$, $K$, $D$, $D_s$, $B$, $B_s$) have been
described  in Ref.~\cite{IS} in which a Gaussian form was used for
the vertex function and free propagators were adopted for the
constituent quarks. The adjustable parameters, the widths of
Bethe-Salpeter amplitudes in momentum space and the constituent
quark masses, were determined from a least square fit to
available experimental data and some lattice determinations. We
found that our results are in good agreement with experimental
data and other approaches. It was also shown that the scaling
relations resulting from the spin-flavor symmetries are reproduced
by the model in the heavy quark limit.

Using this approach  we have elaborated the so-called
{\it Relativistic Three-Quark Model (RTQM)} to study the
properties of heavy baryons containing a single heavy quark
(bottom or charm). For the heavy quarks we used propagators
appropriate for the heavy quark limit. Physical observables for the
semileptonic and nonleptonic decays as well as for the one-pion
and one-photon transitions have been successfully described  in this approach
\cite{RTQM}. Recently, the RTQM was extended to include the
effects of finite quark masses \cite{RTQM1}.
We mention that the authors of \cite{Gatto} have developed
a relativistic quark model approach to the description of  meson
transitions which has similarities to our approach. They also use
an effective heavy meson Lagrangian to describe the couplings of mesons
to quarks. They use, however,  point-like meson-quark interactions.
Loop momenta are explicitly cut off at around 1 GeV in the approach
\cite{Gatto}. In our approach we use momentum dependent meson-quark
interactions which  provides for an effective cut off of the loop integration.
We would also like to mention a recent investigation \cite{NW} where the
same  quark-meson Lagrangian employed in \cite{RCQM} was used.
The authors of  \cite{NW} employed  dipole vertex to describe various
leptonic and semileptonic decays of both the heavy-light mesons and
the $B_c$-meson.

In this paper we follow the strategy adopted in
Refs.~\cite{DSEH,IS}. The basic assumption on the
choice of the vertex function in the hadronic matrix elements is made
after transition to the momentum space. We employ the impulse approximation
in calculating these matrix elements which has been used widely
in phenomenological DSE studies (see,  e.g., Ref.~\cite{DSEH}).
In the impulse approximation one assumes that the vertex functions
depend only on the loop momentum flowing through the vertex.
We present  a general method which greatly facilitates the numerical
evaluations that occur in the Feynman-type calculations involving
quark loops (see also \cite{RCQM,IS}).

The basic emphasis of this work is to study  leptonic and
semileptonic decays of the $B_c$ meson. We use Gaussian
vertex functions with  size parameters  for heavy-light
mesons as in Ref.~\cite{IS}. In this paper we limit our attention
to the basic  semileptonic decay modes of the $B_c$-meson.
A new feature of our calculation is that we also discuss semileptonic decays
involving the $\tau$-lepton. We discuss in some detail how our quark loop
calculations reproduce the heavy quark limit relations between form factors
at zero recoil. Explicit expressions for the reduced set of form factors
in this limit are given.

\section{Model}

We employ an approach \cite{RCQM} based on the effective
interaction Lagrangian which describes the coupling between
hadrons and their constituent quarks. For example, the coupling of
the meson $H$ into its constituents $q_1$ and $
q_2 $ is given by the Lagrangian
\begin{equation}
\label{lag}
{\cal L}_{{\rm int}} (x)=g_H H(x) \int\!\! dx_1 \!\!\int\!\! dx_2
\Phi_H (x,x_1,x_2) \bar q(x_1) \Gamma_H \lambda_H q(x_2)\,.
\end{equation}
Here, $\lambda_H$ and $\Gamma_H$ are  Gell-Mann and Dirac
matrices, respectively, which entail the flavor and spin quantum numbers
of the meson $H$. The function $\Phi_H$ is related to the scalar
part of Bethe-Salpeter amplitude and characterizes the finite size
of the meson.
$\Phi_H$ is invariant under the translation
 $
 \Phi_H(x+a,x_1+a,x_2+a)=\Phi_H(x,x_1,x_2)
 $
which is necessary for the Lorence invariance of the Lagrangian (\ref{lag}).
For instance, the separable form

\begin{equation}
\Phi_H(x,x_1,x_2)=\delta\biggl(x-\frac{x_1+x_2}{2}\biggr)
f((x_1-x_2)^2)
\end{equation}
has been used in \cite{RCQM} for pions with $f(x^2)$ being a Gaussian.
The straightforward generalization of the vertex function (2.2)
to the case of an arbitrary pair of quarks with different masses
is given by
\begin{equation}
\Phi_H(x,x_1,x_2)=
\delta\biggl(x-\frac{m_1x_1+m_2x_2}{m_1+m_2}\biggr) f((x_1-x_2)^2).
\end{equation}
The authors of \cite{NW} used a dipole form for the Fourier-transform of
the function $f(x^2)$. Here we follow the slightly different strategy
as proposed in Refs.~\cite{DSEH,IS}. The choice of the vertex function
in the hadronic matrix elements is specified after transition to momentum
space. We employ the impulse approximation  in calculating the one-loop
transition amplitudes. In the impulse approximation one assumes that
the vertex functions depend only on the loop momentum flowing through
the vertex. The impulse approximation  has been used widely
in phenomenological DSE studies (see,  e.g., Ref.~\cite{DSEH}).
The final results of calculating a quark loop diagram depends on the choice
of loop momentum flow. In the heavy quark transitions discussed in this paper
the loop momentum flow is, however, fixed if one wants to reproduce
the heavy quark symmetry results.

\begin{figure}[t]
\begin{center}
\epsfig{file=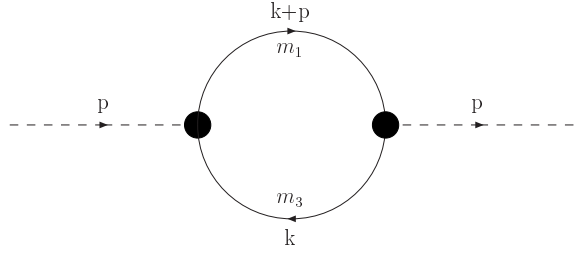,height=4cm}
\caption{One-loop self-energy type diagram needed for the evaluation of
the compositeness condition.}
\label{f:mesonmass}
\end{center}
\end{figure}

 To demonstrate our assumption, we consider the meson mass function
 defined by the diagram in Fig.~\ref{f:mesonmass}. We have

 \begin{eqnarray}
 \label{mass}
 \Pi_H(x-y)&=&\int\!\!dx_1\!\!\int\!\!dx_2 \Phi_H(x,x_1,x_2)
              \int\!\!dy_1\!\!\int\!\!dy_2\Phi_H(y,y_1,y_2)
 \\
 &&\cdot{\rm tr}\left\{S(y_1-x_1)\Gamma_H S(x_2-y_2)\Gamma_H\right\}.
 \nonumber
 \end{eqnarray}
 Then we calculate the Fourier-transform of the meson mass function
 (\ref{mass}).
 \begin{eqnarray}
 &&\tilde\Pi_H(p)=\int e^{-ipx}\Pi_H(x)=\\
 &&\int\!\!\frac{dq}{(2\pi)^4}\!\!
 \int\!\!\frac{dk_1}{(2\pi)^4}\!\!
 \int\!\!\frac{dk_2}{(2\pi)^4}
 \tilde\Phi_H(-p,k_1,-k_2)\tilde\Phi_H(q,-k_1,k_2)
 {\rm tr}\left\{S(\not\! k_1)\Gamma_H S(\not\! k_2)\Gamma_H\right\}.
 \nonumber
 \end{eqnarray}
  The Fourier-transform of the function $\Phi(x_1,...,x_n)$ which
  invariant under the translation $x_i\to x_i+a$ can be written as

 \begin{eqnarray}
  \tilde\Phi(q_1,...,q_n)&=&\int\!\! dx_1\!\!...\int\!\! dx_n
   e^{i\sum\limits_{i=1}^nx_iq_i}\Phi(x_1,...,x_n)\\
 &&
   \cdot(2\pi)^4\delta\left(\sum\limits_{i=1}^n q_i\right)
   \cdot n^4 \int\!\! dx_1\!\!...\int\!\! dx_n
   \delta\left(\sum\limits_{i=1}^n x_i\right)
    e^{i\sum\limits_{i=1}^nx_iq_i}\Phi(x_1,...,x_n)\nonumber\\
  &\equiv& (2\pi)^4\delta\left(\sum\limits_{i=1}^n q_i\right)
  \phi(q_1,...,q_{n-1}).
  \nonumber
  \end{eqnarray}
  Using this property one finds
 \begin{equation}
 \tilde\Pi_H(p)=
  \int\!\!\frac{dk}{(2\pi)^4}\phi_H^2(k,p)
 {\rm tr}\left\{S(\not\! k+\not\! p)\Gamma_H S(\not\! k)\Gamma_H\right\}.
 \end{equation}
  Here, we assume that the vertex function $\phi_H$ depends only on
  the loop momentum $k$. Besides, we assume that $\phi_H$ is analytical
  function which decreases sufficiently fast in the Euclidean momentum space
  to render all loop diagrams UV finite.

The coupling constants $g_H$ is determined  by the so called {\it
compositeness condition} proposed in \cite{SWH} and extensively
used in \cite{EI}. The compositeness condition means that the
renormalization constant of the meson field is equal to zero

\begin{equation}
\label{z=0}
Z_H=1-\frac{3g^2_H}{4\pi^2}\tilde\Pi^\prime_H(m^2_H)=0.
\end{equation}
where $\tilde\Pi^\prime_H$ is the derivative of the meson mass
function defined by the diagram in Fig. \ref{f:mesonmass}

\begin{eqnarray}
\Pi_P(p^2)&=& \int\!\! \frac{d^4k}{4\pi^2i} \phi^2_P(-k^2) {\rm
tr} \biggl[\gamma^5 S_3(\not\! k) \gamma^5 S_1(\not\! k+\not\!
p)\biggr] \,, \label{massp}\\
\Pi_V(p^2)&=&\frac{1}{3}\biggl[g^{\mu\nu}-\frac{p^\mu
p^\nu}{p^2}\biggr] \int\!\! \frac{d^4k}{4\pi^2i} \phi^2_V(-k^2)
{\rm tr} \biggl[\gamma^\mu S_3(\not\! k) \gamma^\nu S_1(\not\!
k+\not\! p)\biggr] \,.\label{massv}
\end{eqnarray}
 For simplicity, we extract the factor $1/4\pi^2$ from the definition
 of the meson mass operator.
We use the local quark propagators

\begin{equation}
S_i(\not\! k)=\frac{1}{m_i-\not\! k} \,,
\end{equation}
where $m_i$ is  the constituent quark mass. As discussed in
\cite{RCQM}, we assume that $m_H<m_{q_1}+m_{q_2}$ in order to
avoid the appearance of imaginary parts in the physical
amplitudes. This is a reliable approximation for the heavy
pseudoscalar mesons. the above condition is not always met for heavy vector
mesons. As discussed in Sec.VI we shall therefore employ equal masses for the
heavy pseudoscalar and vector mesons in our matrix element calculations
but use physical masses for the phase space.

\section{A method for the evaluation of one-loop diagrams with
arbitrary  vertex functions}

For the present purposes one has to evaluate one-loop integrals of
two- and three-point functions involving tensor integrands and product
of vertex functions. In this section we describe a general method
to efficiently enact these calculations for the general case of n-point
one-loop functions. We note two simplifying features of our integration
technique. The arising tensor integrals are reduced to simple invariant
integrations. The sequence of integrations is arranged such that the product
of vertex functions is kept to the very end and allowing for a full
flexibility in the choice of vertex functions.

\begin{figure}[t]
\begin{center}
\epsfig{file=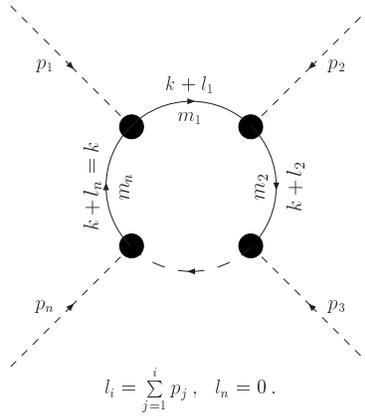,height=6cm}
\caption{One-loop diagram with n-legs and arbitrary vertex functions.}
\label{f:npoints}
\end{center}
\end{figure}

We consider a rank $s$ tensor integral in the Minkowsky  space as
it appears in a general one fermion-loop calculation of a n-point
function (see the diagram in Fig. \ref{f:npoints}). One has

\begin{equation}
I^{\mu_1,...,\mu_s}_{[n,s]}=\int\frac{d^4 k}{i\pi^2}{\cal F}(-k^2)
\frac{k^{\mu_1}...\,k^{\mu_s}}{\prod\limits_{i=1}^n
\left[m^2_i-(k+l_i)^2\right]}
\end{equation}
The outer momenta  $ p_j$ ($j=1,...,n$) are all taken to be
incoming. The momenta of the inner lines are given by $k+l_i$ with
$l_i=\sum\limits_{j=1}^i p_j$ such that $l_n=0$. The  maximum
degree of the momentum tensor  in the numerator arising from the $n$
fermion propagators is  $s_{\rm max}=n$. We employ the impulse
approximation dropping the dependence on the external momenta
inside the vertex functions and denote the product of all vertex
functions  by ${\cal F}(-k^2)$.

Using the $\alpha$-parameterization of  Feynman  one finds

\begin{equation}
I^{\mu_1,...,\mu_s}_{[n,s]}= \Gamma(n)\int
d^n\alpha\delta(1-\sum\limits_{i=1}^n\alpha_i) \int\frac{d^4
k}{i\pi^2}{\cal F}(-k^2) \frac{k^{\mu_1}...\,k^{\mu_s}}
{[D_n(\alpha)-(k+P)^2]^n}\,,
\end{equation}
where $P=\sum\limits_{i=1}^n \alpha_i l_i$, and
$D_n(\alpha)=\sum\limits_{i,j}\alpha_i\alpha_j\, d_{ij}$ with
$d_{i,j}=(1/2)\,[m^2_i+m^2_j-(l_i-l_j)^2]$.

Next we use the Cauchy integral representation for the function
${\cal F}(-k^2)$ leading to

\begin{eqnarray*}
I^{\mu_1,...,\mu_s}_{[n,s]}&=& \Gamma(n)\int
d^n\alpha\delta(1-\sum\limits_{i=1}^n\alpha_i) \int\frac{d^4
k}{i\pi^2} \oint\frac{d\zeta{\cal F}(-\zeta)}{2\pi i}
\frac{k^{\mu_1}...\,k^{\mu_s}} {[\zeta-k^2]
 [D_n(\alpha)-(k+P)^2]^n}\,.
\end{eqnarray*}
The new denominator factor is then included again via Feynman
parameterization giving

\begin{eqnarray*}
I^{\mu_1,...,\mu_s}_{[n,s]}&=& \Gamma(n+1)\int\limits_0^1 d\beta
\beta^{n-1} \int d^n\alpha\delta(1-\sum\limits_{i=1}^n\alpha_i)
\int\frac{d^4 k}{i\pi^2} \oint\frac{d\zeta{\cal F}(-\zeta)}{2\pi
i}
\\
&\times& \frac{k^{\mu_1}...k^{\mu_s}} {
\biggl[(1-\beta)\,\zeta-(k+\beta P)^2+ \beta\,
D_n(\alpha)-\beta(1-\beta)\, P^2 \biggr]^{n+1} }\,.
\end{eqnarray*}
One then factors out the $(1-\beta)$ in the denominator and shift
the integration variable $k$ to $k'=(k+\beta P)/\sqrt{1-\beta}\,$ to
obtain

\begin{eqnarray*}
I^{\mu_1,...,\mu_s}_{[n,s]}&=& \Gamma(n+1)\int\limits_0^1 d\beta
\left(\frac{\beta}{1-\beta}\right)^{n-1} \int
d^n\alpha\delta(1-\sum\limits_{i=1}^n\alpha_i) \int\frac{d^4
k}{i\pi^2} \oint\frac{d\zeta{\cal F}(-\zeta)}{2\pi i}
\\
&\times& \frac{\left(\sqrt{1-\beta}\,k-\beta \,P\right)^{\mu_1}...
\left(\sqrt{1-\beta}\,k-\beta \,P\right)^{\mu_s}}
{[\zeta-k^2+z]^{n+1}}\,.
\end{eqnarray*}

The contour integral can be done again by Cauchy's theorem.
On substitution of $\beta=t/(1+t)$ one then has

\begin{eqnarray*}
I^{\mu_1,...,\mu_s}_{[n,s]}&=& (-)^n\int\limits_0^\infty dt\frac{
t^{n-1}}{(1+t)^2} \int
d^n\alpha\delta(1-\sum\limits_{i=1}^n\alpha_i) \int\frac{d^4
k}{i\pi^2}{\cal F}^{(n)}(-k^2+z)\, K^{\mu_1}...K^{\mu_s}\,,
\end{eqnarray*}
where ${\cal F}^{(n)}$ denotes the $n$-th derivative of the function
${\cal F}$ and where

$$ K^\mu=\frac{1}{\sqrt{1+t}}\,k^\mu-\frac{t}{1+t}\, P^\mu,
\hspace{1cm} z=t\,D_n(\alpha)-\frac{t}{1+t}\,P^2. $$ The momentum
integration of the tensor integral can be trivially done by
invariant integration. Finally we go to the Euclidean space by
rotating $k_0\to ik_4$ which gives  $k^2\to-k^2_E\equiv u$,
then one encounters the scalar integrals

\begin{equation}
I_{[n,m]}= (-)^n\int\limits_0^\infty dt \frac{
t^{n-1}}{(1+t)^{2+m}} \int
d^n\alpha\delta(1-\sum\limits_{i=1}^n\alpha_i)
\int\limits_0^\infty du u^{m+1} {\cal F}^{(n)}(u+z)\,.
\end{equation}
The u-integration can be performed by partial integration and one
finally obtains

\begin{eqnarray}
\label{loop} I_{[n,m]}= (-)^{n+m}\Gamma(m+2)\int\limits_0^\infty
dt\frac{ t^{n-1}}{(1+t)^{2+m}} \int
d^n\alpha\delta(1-\sum\limits_{i=1}^n\alpha_i) {\cal
F}^{(n-m-2)}(z)\,.
\end{eqnarray}
Since $m_{\rm max}=[n/2]$ Eq.~(\ref{loop}) holds true for  all n and m
except for the case $n=2$ and $m=1$. In this case we have

\begin{eqnarray}
I_{[2,1]}&=& 2\int\limits_0^\infty dt\frac{t}{(1+t)^3} \int
d^2\alpha\delta(1-\sum\limits_{i=1}^2\alpha_i)
\int\limits_z^\infty du {\cal F}(u) \nonumber\\ &=&
\int\limits_0^\infty dt \left(\frac{t}{1+t}\right)^2 \int
d^2\alpha\delta(1-\sum\limits_{i=1}^2\alpha_i) z'_t{\cal F}(z)\,.
\end{eqnarray}
where $z'_t=dz(t)/dt$.

One has to remark that the integration over the
$\alpha-$parameters in Eq.~(\ref{loop}) can be done analytically
up to a remaining  one-fold integral. However, the ease with which
the numerical $\alpha$-integrations can be done does not warrant
the effort of further analytical integrations. Using the integration
techniques described in this section all necessary numerical integrations
encountered in this investigation can be performed within minutes using
a fast modern PC's.

\section{Hadronic matrix elements}

\subsection{Quark-meson coupling constants}

As already discussed in Sec.II, the quark-meson coupling constants
are determined by the compositeness condition Eq.~(\ref{z=0}).
The derivatives of the meson-mass functions  can be written as

\begin{eqnarray}
\frac{d}{dp^2}\Pi_P(p^2)&=&\frac{1}{2p^2}p^\alpha\frac{d}{dp^\alpha}
\int\!\! \frac{d^4k}{4\pi^2i} \phi^2_P(-k^2) {\rm tr}
\biggl[\gamma^5 S_3(\not\! k) \gamma^5 S_1(\not\! k+\not\!
p)\biggr] \, \nonumber\\ &=& \frac{1}{2p^2}\int\!\!
\frac{d^4k}{4\pi^2i} \phi^2_P(-k^2) {\rm tr} \biggl[\gamma^5
S_3(\not\! k) \gamma^5 S_1(\not\! k+\not\! p)\not\! p \,S_1(\not\!
k+\not\! p)\biggr] \,,
\\
&&\nonumber\\ \frac{d}{dp^2}\Pi_V(p^2)&=&\frac{1}{3}
\frac{1}{2p^2}p^\alpha\frac{d}{dp^\alpha}
\biggl[g^{\mu\nu}-\frac{p^\mu p^\nu}{p^2}\biggr] \int\!\!
\frac{d^4k}{4\pi^2i} \phi^2_V(-k^2) {\rm tr} \biggl[\gamma^\mu
S_3(\not\! k) \gamma^\nu S_1(\not\! k+\not\! p)\biggr] \,
\\
&=& \frac{1}{3} \biggl[g^{\mu\nu}-\frac{p^\mu p^\nu}{p^2}\biggr]
\frac{1}{2p^2}\int\!\! \frac{d^4k}{4\pi^2i} \phi^2_V(-k^2) {\rm
tr} \biggl[\gamma^\mu S_3(\not\! k) \gamma^\nu S_1(\not\! k+\not\!
p)\not\! p\,S_1(\not\! k+\not\! p)\biggr] \,. \nonumber
\end{eqnarray}

The evaluation of the integrals is done by using the method
outlined in Sec.III. The compositeness condition reads

\begin{equation}
\label{norm}
\frac{3g^2_H}{4\pi^2}N_H=1, \hspace{0.5cm} {\rm where}
\hspace{0.5cm} N_H=\frac{d}{dp^2}\Pi_H(p^2)|_{p^2=m^2_H}
\end{equation}

\begin{eqnarray*}
N_H&=& \frac{1}{2}\int\limits_0^\infty dt
\biggl(\frac{t}{1+t}\biggr)^2 \int\limits_0^1 d\alpha\, \alpha
\biggl\{...\biggr\}_H
\\
&&\\ \biggl\{...\biggr\}_P &=& {\cal F}_P(z)\, \frac{1}{1+t}
\biggl[4-3\frac{\alpha t}{1+t}\biggr] -{\cal F^{\prime}}_{\!\!P}(z)
\biggl\{ 2m_1m_3 +\frac{\alpha
t}{1+t}\biggl[m_1^2-2m_1m_3+p^2\biggr]
\\
&& -p^2\biggl(\frac{\alpha t}{1+t}\biggr)^2 \left(2-\frac{\alpha
t}{1+t}\right) \biggr\}\,,
\\
&&\\ \biggl\{...\biggr\}_V &=&{\cal F}_V(z)  \frac{1}{1+t}
\biggl[2-\frac{\alpha t}{1+t}\biggr] -{\cal F^{\prime}}_{\!\!V}(z)
\biggl\{ 2m_1m_3 +\frac{\alpha
t}{1+t}\biggl[m_1^2-2m_1m_3+p^2\biggr]
\\
&& -p^2\biggl(\frac{\alpha t}{1+t}\biggr)^2 \left(2-\frac{\alpha
t}{1+t}\right) \biggr\}\,.
\end{eqnarray*}
Here, $m_1$ stands for the heavy quark ($b$ or $c$) and $m_2$ for
the light quarks ($u$, $d$, $s$) in the case of heavy-light
systems, and for $c$ in the case of double-heavy systems. The
function ${\cal F}_H(z)$ is the product of two vertex functions
${\cal F}_H(z)=\phi_H^2(z)$ with

$$ z=t\,\left(\alpha m^2_1+(1-\alpha) m^2_3-\alpha(1-\alpha) p^2\right)
-\frac{\alpha^2 t}{1+t}\, p^2. $$

\subsection{Leptonic and radiative decays}

\begin{figure}[t]
\begin{center}
\epsfig{file=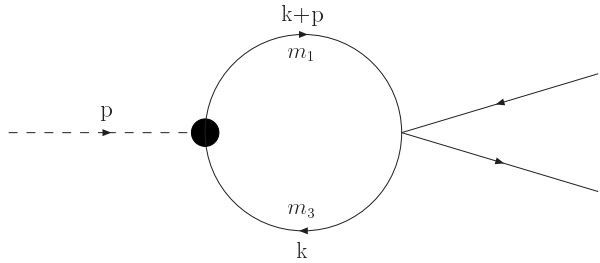,height=4cm}
\caption{Quark model diagram for leptonic meson decays.}
\label{f:leptonicdecay}
\end{center}
\end{figure}

\begin{figure}[t]
\begin{center}
\epsfig{file=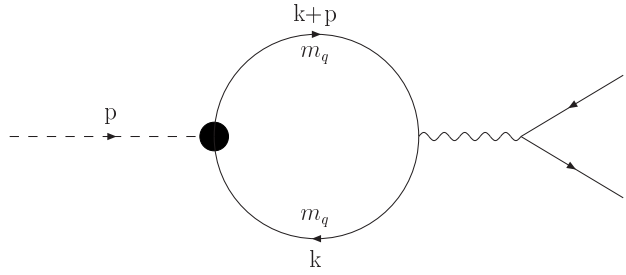,height=4cm}
\caption{Quark model diagram for vector meson radiative decays.}
\label{f:radiative}
\end{center}
\end{figure}

\begin{figure}[t]
\begin{center}
\epsfig{file=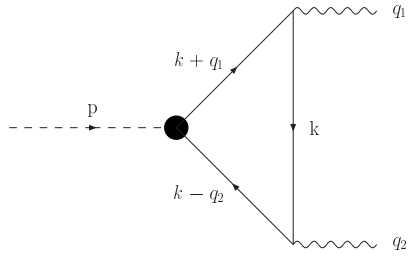,height=4cm}
\caption{Quark model diagram for  the decays of a neutral meson.}
\label{f:PGG}
\end{center}
\end{figure}

\begin{figure}[t]
\begin{center}
\epsfig{file=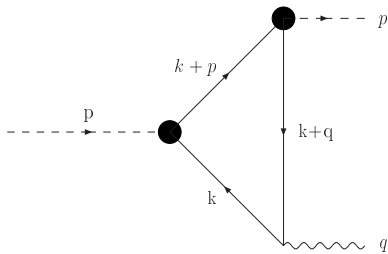,height=4cm}
\caption{Quark model diagram for  the radiative decay of a vector meson
into a pseudoscalar one.}
\label{f:VPG}
\end{center}
\end{figure}

The matrix elements of the leptonic and radiative decays are defined
by the diagrams in Figs.~\ref{f:leptonicdecay}-\ref{f:VPG} and given by

\begin{eqnarray}
iM^\mu_P(p)& = &
\frac{3g_P}{4\pi^2}\!\!\int\!\!\frac{d^4k}{4\pi^2i} \phi_P(-k^2)
{\rm tr}\biggl[\gamma^5 S_3(\not\! k) O^\mu S_1(\not\! k+\not\! p)
\biggr] =f_P p^\mu\,,
\\
&&\nonumber\\
M^\mu_V(p)& = -&C_V\cdot
\frac{3g_V}{4\pi^2}\!\!\int\!\!\frac{d^4k}{4\pi^2i} \phi_V(-k^2)
{\rm tr}\biggl[\not\! \epsilon^\ast S_q(\not\! k) \gamma^\mu
S_q(\not\! k+\not\! p) \biggr]
\nonumber\\
&=&-m_V C_V f_V
\epsilon^{\ast\mu}\,,
\\
&&\nonumber\\
iM_{P\gamma\gamma}(q_1,q_2)& = &
C_{P\gamma\gamma}\frac{3g_P}{4\pi^2}\!\!\int\!\!\frac{d^4k}{4\pi^2i}
\phi_P(-k^2) {\rm tr}\biggl[\gamma^5 S_q(\not\! k-\not\! q_2)
\not\! \epsilon^\ast_2 S_q(\not\! k)\not\! \epsilon^\ast_1
S_q(\not\! k+\not\! q_1) \biggr]\,,
\nonumber\\
&=&ig_{P\gamma\gamma}\,
\varepsilon^{\mu\nu\alpha\beta}\epsilon_1^{\ast\mu}\epsilon_2^{\ast\nu}
q_1^\alpha q_2^\beta\,,
\\
&&\nonumber\\
iM_{VP\gamma}(p,p')& = &
C_{VP\gamma}\frac{3g_Vg_P}{4\pi^2}\!\!\int\!\!\frac{d^4k}{4\pi^2i}
{\cal F}_{PV}(-k^2)\, {\rm tr}\biggl[\gamma^5 S_q(\not\! k+\not\!
p) \not\! \epsilon^\ast  S_q(\not\! k)\not\!
\epsilon^\ast_\gamma S_q(\not\! k+\not\! q) \biggr] \nonumber\\
&=&ig_{VP\gamma}\,\varepsilon^{\mu\nu\alpha\beta}
\epsilon_\gamma^{\ast\mu}\epsilon^{\ast\nu} p^\alpha q^\beta\,.
\end{eqnarray}
For ease of presentation, the expression for the $V\to P\gamma$-decay is given
for neutral-flavored mesons. Using the integration techniques described
in Sec.III one then arrives at the following analytical representation
of the various one-loop matrix elements

\begin{eqnarray}
f_P&=& \frac{3g_P}{4\pi^2}\int\limits_0^\infty dt
\frac{t}{(1+t)^2} \int\limits_0^1 d\alpha \phi_P(z_P) \biggl[m_3
+(m_1-m_3)\frac{\alpha t}{1+t}\biggr]\,,\\
&&z_P=t\,\left(\alpha
m_1^2+(1-\alpha) m_3^2-\alpha(1-\alpha)p^2\right) -
\frac{t\alpha^2}{1+t}\, p^2 \,.\nonumber\\
&&\nonumber\\ f_V&=&
\frac{1}{m_V}\frac{3g_V}{4\pi^2} \int\limits_0^\infty dt
\frac{t}{(1+t)^2} \int\limits_0^1 d\alpha \phi_V(z_V)
\biggl[m_q^2+\frac{1}{2}tz^{\prime}_t + \frac{\alpha t}{1+t}\,
\left(1 -\frac{\alpha t}{1+t}\right)\, p^2 \biggr]\,,
\\
&&z_V=t\,\left(m^2_q-\alpha(1-\alpha)p^2\right) - \frac{t\alpha^2}{1+t} \,
p^2 \,, \nonumber\\ &&\Gamma(V\to
e^+e^-)=\frac{4\pi}{3}\,\frac{\alpha^2}{m_V}\, f^2_V \,C_V^2\,,
\hspace{1cm} C_V=e^2_q\,\,\,\, (V=\phi\,\,,J/\psi\,\,,\Upsilon).
\nonumber\\ \nonumber&&\\
g_{P\gamma\gamma}&=&C_{P\gamma\gamma}\cdot m_q \cdot
\frac{3g_P}{4\pi^2} \int\limits_0^\infty dt
\biggl(\frac{t}{1+t}\biggr)^2 \int d^3\alpha
\delta\biggl(1-\sum\limits_{i=1}^3\alpha_i\biggr)
\biggl(-\phi_P^{\prime}(z_0)\biggr)\,,
\\
&&z_0=t\,\left(m^2_q-\alpha_1\alpha_2 p^2\right) -\frac{t}{1+t}
\alpha_1\alpha_2 p^2 \,, \nonumber\\
&&\Gamma(P\to\gamma\gamma)=
\frac{\pi}{4}\alpha^2 m^3_Pg^2_{P\gamma\gamma}\,, \hspace{1cm}
C_{\eta_c\gamma\gamma}=2e_c^2\,, \nonumber\\
&&\nonumber\\
g_{VP\gamma}&=&C_{VP\gamma}\cdot m_q \cdot \frac{3g_Vg_P}{4\pi^2}
\int\limits_0^\infty dt \biggl(\frac{t}{1+t}\biggr)^2 \int
d^3\alpha \delta\biggl(1-\sum\limits_{i=1}^3\alpha_i\biggr)
\biggl({-\cal F'}_{VP}(z_{VP})\biggr)\,,
\\
&&z_{VP}= t\,\left(m^2_q-\alpha_1\alpha_3\, m^2_V-\alpha_1\alpha_2\,
m^2_P\right) -\frac{t}{1+t} \,\left(\alpha_1(\alpha_1+\alpha_2) m^2_V
-\alpha_1\alpha_2 m^2_P)\right) \,, \nonumber\\ &&\Gamma(V\to P\gamma)=
\frac{\alpha}{24} m^3_V \biggl(1-\frac{m^2_P}{m^2_V}\biggr)^3
g^2_{VP\gamma}\, \hspace{1cm} C_{J/\psi\eta_c\gamma}=2e_c\,.
\nonumber
\end{eqnarray}
The electric quark charges $e_q$ are given in units of $e$.

\subsection{Semileptonic form factors}

\begin{figure}[t]
\begin{center}
\epsfig{file=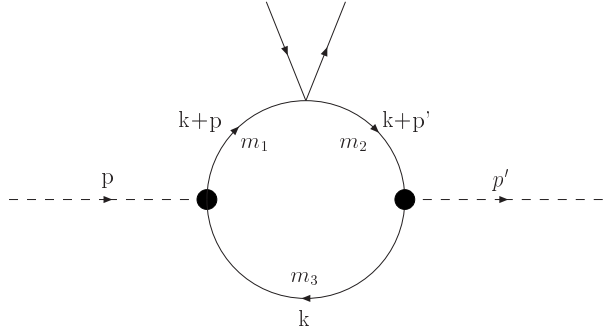,height=5cm}
\caption{Quark model diagram for the semileptonic $B_c$-decays involving
$b\to c,u$ transitions. Lower leg in loop is c-quark.}
\label{f:bdecay}
\end{center}
\end{figure}

The semileptonic decays of the $B_c$-meson can be induced by either
a beauty quark or a charm quark transition. In the relativistic quark
model, the hadronic matrix element corresponding to b-decay is
defined by the diagram in Fig.~\ref{f:bdecay} and is given by

\begin{equation}
\label{b-dec}
M^\mu_b(P(p)\to H(p')) =
\frac{3g_Pg_H}{4\pi^2}\!\!\int\!\!\frac{d^4k}{4\pi^2i} {\cal
F}_{PH}(-k^2) {\rm tr}\biggl[\gamma^5 S_3(\not\! k) \Gamma_H
S_2(\not\! k+\not\! p') O^\mu S_1(\not\! k+\not\! p) \biggr]\,,
\end{equation}
where ${\cal F}_{PH}=\phi_P\cdot\phi_H$, $\Gamma_P=\gamma^5$, and
$\Gamma_V=-i\not\!\epsilon^\ast$ with $\epsilon^\ast\cdot p'=0$.
For the b-decay case one has the CKM-enhanced decays
\begin{eqnarray*}
&&b\to c: \hspace{0.5cm} B_c^+ \to  (\eta_c,J/\psi)\, l^+\nu
\hspace{0.5cm}\, m_1 =m_b\,,\,\,m_2=m_3=m_c\,,
\end{eqnarray*}
and the CKM-suppressed decays
\begin{eqnarray*}
&&b\to u: \hspace{0.5cm} B_c^+ \to  (D^0,D^{\ast 0})\, l^+\nu
\hspace{0.5cm} m_1 =m_b\,,\,\,m_2=m_u\,,\,\,m_3=m_c\,.
\end{eqnarray*}

\begin{figure}[t]
\begin{center}
\epsfig{file=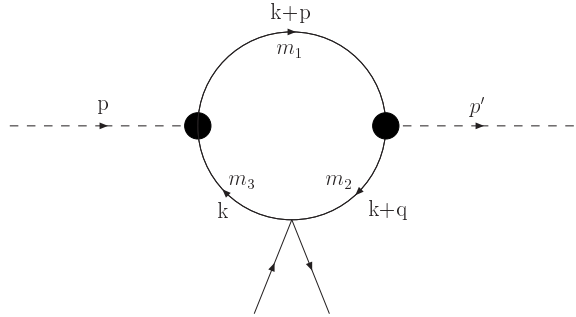,height=5cm}
\caption{Quark model diagram for the semileptonic $B_c$-decays involving
$c\to s,d$ transitions. Upper leg in loop is b-quark. }
\label{f:cdecay}
\end{center}
\end{figure}

The c-decay option of the $B_c$-meson is represented  by the Feynman
diagram in Fig.~\ref{f:cdecay} which gives

\begin{equation}
\label{c-dec} M^\mu_c(P(p)\to H(p'))= \frac{3g_Pg_H}{4\pi^2}\!\!
\int\!\!  \frac{d^4k}{4\pi^2i} {\cal F}_{PH}(-k^2) {\rm
tr}\biggl[\gamma^5 S_3(\not\! k) O^\mu S_2(\not\! k+\not\! q)
\Gamma_H S_1(\not\! k+\not\! p) \biggr] \,.
\end{equation}
Again one has the CKM-enhanced decays

\begin{eqnarray*}
&&c\to s: \hspace{0.5cm} B_c^+ \to (\bar B^0_s,\bar B^{\ast
0}_s)\, l^+\nu  \hspace{0.5cm} m_1 =m_b\,,\,\,
m_2=m_s\,,\,\,m_3=m_c.
\end{eqnarray*}
and the CKM-suppressed decays
\begin{eqnarray*}
&&c\to d: \hspace{0.5cm} B_c^+ \to  (\bar B^0,\bar  B^{\ast 0})
\,l^+\nu  \hspace{0.5cm} m_1 =m_b\,,\,\, m_2=m_d \,,\,\,
m_3=m_c\,.
\end{eqnarray*}
It is convenient to present the results of the matrix element
evaluations in terms of invariant form factors.
A standard decomposition of the transition matrix elements into
invariant form factors is given by

\begin{eqnarray}
M^\mu(P(p)\to P'(p'))&=&f_+(q^2)\;(p+p^{\prime})^\mu\,+\,
f_-(q^2)\;(p-p^{\prime})^\mu \label{PP}
\end{eqnarray}
and

\begin{eqnarray}
iM^\mu(P(p)\to V(p')) &=& -g^{\mu\nu}\epsilon^{\ast\nu}
(m_{P}+m_{V})\,A_1(q^2)+ (p+p')^\mu\,  p\cdot \epsilon^\ast \,
\frac{A_2(q^2)}{m_P+m_V} \label{PV}
\\
&& +(p-p')^\mu\, p\cdot \epsilon^\ast \,\frac{A_3(q^2)}{m_P+m_V}
-i\varepsilon ^{\mu\nu\alpha\beta}\epsilon^{\ast\nu} p^\alpha
p^{\prime~\beta }\, \frac{2\ V(q^2)}{m_P+m_V}\,. \nonumber
\end{eqnarray}

The various invariant form factors can be extracted from the one-loop
expressions (\ref{b-dec}) and (\ref{c-dec}) by using the techniques described
in Sec.III. One finds that the form factor integrands factorize into
a common piece times a piece specific to the different form factors.
One can thus write

\begin{equation}
\label{ff}
F(q^2)=\frac{3}{4\pi^2}\, g_{P}g_{H}\,
\frac{1}{2}\int\limits_0^\infty dt \biggl(\frac{t}{1+t}\biggr)^2
\int\!
d^3\alpha\,\delta\biggl(1-\sum\limits_{i=1}^3\alpha_i\biggr)
\biggl\{...\biggr\}_{F}
\end{equation}
where $F=f_{\pm},A_i,V$. For the $0^-\to 0^-$ $b\to c,u$ form factors
$f_+$ one has

\begin{eqnarray*}
\biggl\{...\biggr\}^b_{f_+}&=& {\cal F}_{PP}(z_b) \frac{1}{1+t}
\biggl[4-3(\alpha_1+\alpha_2)\frac{t}{1+t}\biggr]
 -{\cal F}'_{PP}(z_b) \biggl\{ (m_1+m_2)m_3
\\
&& +\frac{t}{1+t} \biggl(
-(\alpha_1+\alpha_2)(m_1m_3+m_2m_3-m_1m_2) +\alpha_1
\,p^2+\alpha_2\, p'^2 \biggr) \\ && -\biggl(\frac{t}{1+t}\biggr)^2
\biggl( 2-(\alpha_1+\alpha_2)\frac{t}{1+t}\biggr)
\biggl((\alpha_1+\alpha_2)(\alpha_1\, p^2 +\alpha_2 \,p'^2)
-\alpha_1\alpha_2 \,q^2\biggr)\biggr\}\,,
 \\
&& z_b=t\,\left(\sum\limits_{i=1}^3 \alpha_i m^2_i-\alpha_1\alpha_3 p^2
-\alpha_2\alpha_3 p'^2-\alpha_1\alpha_2 q^2\right)-\frac{t}{1+t}\,
P_b^2\,, \hspace{0.5cm} P_b=\alpha_1 p+\alpha_2 p' \,.
\end{eqnarray*}
For the corresponding $c\to s,d$ form factor one has
\begin{eqnarray*}
\biggl\{...\biggr\}^c_{f_+}&=& -{\cal F}_{PP}(z_c)
\frac{1}{1+t} \biggl[1+3\alpha_1\frac{t}{1+t}\biggr]
 +{\cal F}'_{PP}(z_c)
\biggl\{ m_2 m_3
\\
&& +\frac{t}{1+t} \biggl( \alpha_1
(m_1m_2+m_1m_3-m_2m_3)+\alpha_2\, q^2 \biggr) +(\frac{t}{1+t})^2
\biggl(\alpha_1^2\, p^2-\alpha_2^2\, q^2)\biggr)
\\
&& -\left(\frac{t}{1+t}\right)^3
\alpha_1\biggl(\alpha_1(\alpha_1+\alpha_2)\,p^2
-\alpha_1\alpha_2\, p'^2 +\alpha_2 (\alpha_1+\alpha_2)\,
q^2\biggr)\,,
\\
&& z_c=t\,\left(\sum\limits_{i=1}^3 \alpha_i m^2_i-\alpha_1\alpha_3\,
p^2 -\alpha_2\alpha_3\, q^2-\alpha_1\alpha_2\,
p'^2\right)-\frac{t}{1+t}\, P_c^2\,, \hspace{0.5cm} P_c=\alpha_1
p+\alpha_2 q\, .
\end{eqnarray*}
Expressions for the remaining $0^-\to 0^-$ and $0^-\to 1^-$ form factors
$f_-$, $A_i$ and $V$ are given in the Appendix. The masses $m_i\, (i=1,2,3)$
appearing in the form factor expressions are constituent quark masses
with a labelling according to Eqs.~(\ref{b-dec}) and (\ref{c-dec}).
The values of the constituent quark masses as well as the vertex functions
entering the form factor expressions will be specified in Sec.VI.

For the calculation of physical quantities it is more convenient to use
helicity amplitudes. They are linearly related to the invariant form factors
\cite{KS}.
For the $0^-\to 0^-$ transitions one has

\begin{eqnarray}
H_0(q^2)&=&\frac{2 m_P P}{\sqrt{q^2}}\,f_+(q^2)\,,
\\
H_t(q^2)&=&\frac{1}{\sqrt{q^2}} \biggl\{
(m_P^2-m_{P'}^2)\,f_+(q^2)+q^2\,f_-(q^2)\biggl\} \,.
\end{eqnarray}

\noindent
For the $0^-\to 1^-$ transitions one has

\begin{eqnarray}
H_{\pm}(q^2)&=&-(m_P+m_V) A_1(q^2)\ \mp\ \frac{2m_P P}{(m_P+m_V)}\
V(q^2)\,,
\\
H_0(q^2)&=&\frac{1}{2m_V\sqrt{q^2}} \biggl\{
-(m_P^2-m_V^2-q^2)(m_P+m_V)A_1(q^2) +\frac{4m_P^2 P^2}{m_P+m_V}
A_2(q^2)\ \biggr \}\,,
\\
H_t(q^2)&=&\frac{m_P P}{m_V\sqrt{q^2}} \biggl\{ -(m_P+m_V)\,A_1(q^2)
+(m_P-m_V)\, A_2(q^2)+\frac{q^2}{m_P+m_V}\, A_3(q^2)\ \biggr \}\,,
\end{eqnarray}
where
$$ P=\frac{\sqrt{\lambda(m_P^2,m_H^2,q^2)}}{2m_P}\ =\
\frac{[(q^2_+-q^2)(q^2_--q^2)]^{1/2}}{2m_P}
$$
with  $q^2_{\pm}=(m_P \ \pm\ m_H)^2$.

Then the partial helicity rates are defined as

\begin{eqnarray}
\frac{d\Gamma_i}{dq^2}&=& \frac{G^2_F}{(2\pi)^3} \,|V_{ff'}|^2 \cdot
\frac{(q^2-m^2_l)^2\,P}{12m_P^2 q^2}\cdot |H_i(q^2)|^2 \,, \hspace{1cm}
i=\pm,0,t,
\end{eqnarray}
where $V_{ff^{\prime}}$ is the relevant element of the CKM matrix,
$m_l$ is the mass of charged lepton.

Finally, the total partial rates including lepton mass effects can be
written as \cite{KS}

\begin{eqnarray}
\frac{d\Gamma^{\rm PP'}}{dq^2}&=&
(1+\frac{m^2_l}{2q^2})\,\frac{d\Gamma^{\rm PP'}_0}{dq^2} +3\,
\frac{m^2_l}{2q^2}\,\frac{d\Gamma^{\rm PP'}_t}{dq^2}\,,
\\
&&\nonumber\\
\frac{d\Gamma^{\rm PV}}{dq^2}&=& (1+\frac{m^2_l}{2q^2})\,
\biggl[\frac{d\Gamma^{\rm PV}_+}{dq^2}+\frac{d\Gamma^{\rm
PV}_-}{dq^2}+ \frac{d\Gamma^{\rm PV}_0}{dq^2}\biggr] +3\,
\frac{m^2_l}{2q^2}\,\frac{d\Gamma^{\rm PV}_t}{dq^2}\,.
\end{eqnarray}
In the following we shall present numerical results of the total
decay widths, polarization ratio and forward-backward asymmetry.
The relevant expressions are given by

\begin{eqnarray}
\label{width}
\Gamma &=&
\int\limits_{m_l^2}^{(m_P-m_H)^2}dq^2\,\frac{d\Gamma}{dq^2}\,,
\hspace{1cm} \alpha = 2\ \frac{\Gamma_0}{\Gamma_+ +
\Gamma_-}-1\,, \hspace{1cm} A_{FB}=\frac{3}{4}\ \frac{\Gamma_-
- \Gamma_+}{\Gamma}\,.
\end{eqnarray}

\section{Heavy quark spin symmetry}

Our model allows us to evaluate form factors directly from
Eq.~(\ref{ff}) without any approximation. However, it would be
interesting to explore whether the heavy quark spin symmetry relations
derived in Ref.~\cite{Jenkins} can be reproduced in our
approach. As  was shown (see, for instance, \cite{IS}) our model
exhibits all consequences of the spin-flavor symmetry for the
heavy-light systems in the heavy quark limit. For example, the quark-meson
coupling and leptonic decay constants behave as

\begin{eqnarray}
\label{hql1}
g_H&\to& \sqrt{2m_1} \cdot \frac{2\pi}{\sqrt{3\tilde N_H}}\,,
\hspace{1cm} \tilde N_H = \int\limits_0^\infty
du\phi^2_H(u-2E\sqrt{u})\,
\frac{m_3+\sqrt{u}}{m_3^2+u-2E\sqrt{u}}\,,
\\
&&\nonumber\\
\label{hql2}
f_H &\to &
\frac{1}{\sqrt{m_1}}\cdot\sqrt{\frac{3}{2\pi^2\tilde N_H}}
\,\int\limits_0^\infty du[\sqrt{u}-E]\phi_H(u-2E\sqrt{u})\,
\frac{m_3+\sqrt{u}/2}{m_3^2+u-2E\sqrt{u}}\,,
\end{eqnarray}
in the heavy quark limit: $p^2=m_H^2=(m_1+E)^2$ when $m_1\to \infty$.
Eqs.~(\ref{hql1}) and (\ref{hql2}) make the heavy quark mass dependence
of the coupling factors $g_H$ and $f_H$ explicit since we have factorized
the coupling factor contributions into a heavy mass dependent  piece and
a remaining heavy mass independent piece. Moreover, Eqs.~(\ref{hql1}) and (\ref{hql2})
show  $g_H$ and $f_H$ scale as $m_1^{1/2}$ and $m_1^{-1/2}$, respectively.

As is well known (see Ref.~\cite{Jenkins}), heavy flavor symmetry
cannot be used for hadrons containing two heavy quarks. But one
can still derive relations near zero recoil by using heavy quark
spin symmetry.

First, we consider the semileptonic decays $B_c\to \bar B_s(\bar
B^0)e^+\nu$ and $B_c\to\bar B_s^*(\bar B^{*0})e^+\nu$  which
correspond to c-decay into light $s$ and $d$ quark, respectively.
Since the energy released in such decays is much less than the
mass of the b-quark  the four-velocity of the $B_c$-meson is almost
unaffected. Then the initial and final meson momenta can be
written as

$$ p=m_{B_c}v \hspace{1cm} p'=m_B v+r $$ where $r$ is a small
residual momentum ($v\cdot r=-r^2/(2m_B)$). The heavy quark spin
symmetry can be realized in the following way. We split the
B-meson masses into the sum of b-quark mass and binding energy

$$ m_{B_c}\equiv m_P=m_1+E_1, \hspace{1cm} m_{B}\equiv
m_H=m_1+E_2.
$$
Then we go to  the heavy quark mass limit
$m_b\equiv m_1\to\infty$ in which  the b-quark propagator acquires the form

\begin{equation}
\label{b-prop}
\frac{1}{m_1-\not\! p-\not\! k}\Longrightarrow
\frac{1+\not\! v}{-2(kv+E_1)}.
\end{equation}
The decoupling of the c-quark spin allows us to reliably neglect
the  $k$-integration because $k$ is small compare to the heavy  c-quark mass.
One has

\begin{equation}
\label{c-prop} \frac{1}{m_3-\not\! k}\Longrightarrow\frac{1}{m_3}.
\end{equation}
As a consequence, the hadronic matrix element describing the weak
c-quark decay simplifies:

\begin{equation}
M^\mu_c= \frac{\sqrt{2m_P\cdot 2 m_H}}{\sqrt{\tilde N_P\cdot
\tilde N_H}} \cdot\frac{1}{m_3}\cdot \int\!\!
\frac{d^4k}{4\pi^2i} {\cal F}_{PH}(-k^2) \frac{{\rm
tr}[O^\mu(m_2+\not\! k+\not\! q)\Gamma_H (1+\not\! v)\gamma^5]}
{[-2kv-2E_1][m_2^2-(k+q)^2]}\,,
\end{equation}
where $q=p-p'=(m_P-m_H)v-r$=$((E_1-E_2)v-r \equiv \Delta E v-r$ and
$m_2$ stands for the light  quark mass ($m_2=m_s$ or $m_d$).
One has to emphasize that all above approximations
are valid only close to the zero-recoil point $q^2_{\rm max}=\Delta E^2$.
Recalling the transversality of the final vector meson field
$p'\cdot \epsilon^\ast$=$m_B v\cdot+r\cdot \epsilon^\ast=0$
and applying the integrations as described  in  Sec.III, one finds

\begin{equation}
\label{c-hql}
M^\mu_c= \frac {\sqrt{2m_P\cdot 2 m_H}} {\sqrt{\tilde N_P\cdot
\tilde N_H}} \cdot\frac{1}{m_3}\cdot \int\limits_0^\infty\!
\frac{dt\, t}{(1+t)^2}\int\limits_0^\infty d\alpha {\cal
F}_{PH}(z_c)\{...\}_{PH}\,,
\end{equation}

\begin{eqnarray*}
\{...\}_{PP}&=& -\biggl(m_2+\frac{\alpha t-\Delta E}{1+t}\biggr)v^\mu
-\frac{1}{1+t}r^\mu\,,
\\
i\{...\}_{PV}&=& -i\varepsilon^{\mu\nu\alpha\beta}\epsilon^{\ast^\nu}
v^\alpha r^\beta \frac{1}{1+t}+\biggl(m_2+\frac{\alpha t-\Delta
E}{1+t}\biggr)\,\epsilon^{\ast\mu} -\frac{1}{1+t}\, v^\mu
\,\epsilon^\ast \cdot r \,.
\end{eqnarray*}
Here,
$
z_c=(\alpha t^2/(1+t)) (\alpha+2\Delta E)+t\, (m_2^2-2\alpha E_1)
-(t/(1+t))\,\Delta E^2.
$
It is readily seen that the amplitudes of  c-decay in the heavy quark limit
are expressed through two independent functions
$$
\{...\}_1=\biggl(m_2+\frac{\alpha t-\Delta E}{1+t}\biggr),
\hspace{0.5cm}
\{...\}_2=\frac{1}{1+t}.
$$

To complete the description of the heavy quark limit in the
c-decay modes, we give the expressions for the form factors in
this limit. One has

\begin{eqnarray}
F(q^2_{\rm max})&\to& \frac {\sqrt{2m_P\cdot 2 m_H}} {\sqrt{\tilde
N_P\cdot \tilde N_H}} \cdot\int\limits_0^\infty\!
\frac{dt\, t}{(1+t)^2}\int\limits_0^\infty d\alpha {\cal
F}_{PH}(z_c)\{...\}_{F}
\end{eqnarray}
where $F=f_{\pm},A_i,V$. The form factor specific pieces are given by

\begin{eqnarray*}
\{...\}_{f_+}&=&-\frac{1}{2m_1m_3}\cdot \biggl(m_2+\frac{\alpha
t}{1+t}\biggr)\,, \hspace{1cm}
\{...\}_{f_-}=\frac{1}{m_3}\cdot\frac{1}{1+t}\,,
\\
\{...\}_{A_1}&=&-\frac{1}{m_P+m_V}\cdot\frac{1}{m_3}\cdot
\biggl(m_2+\frac{\alpha t-\Delta E}{1+t}\biggr)\,,
\\
\{...\}_{A_2}&=&\{...\}_{V}=\frac{m_P+m_V}{2}\cdot
\frac{1}{m_1m_3}\cdot\frac{1}{1+t}\,,
\\
\{...\}_{A_3}&=&=\frac{m_P+m_V}{2}\cdot \frac{1}{m_1m_3}\cdot
\biggl[-\frac{3}{1+t}+4\,(\frac{m_1}{m_3}-1)\cdot\frac{t}{1+t}\biggr]\,.
\end{eqnarray*}
Superficially it appears that the form factors $f_+$ and $A_1$ are
suppressed by a factor of $1/m_1$. However, they must be kept in
the full amplitude to obtain the correct result in
Eq.~(\ref{c-hql}), for instance, one has

$$ f_+(p+p')^\mu+f_-(p-p')^\mu= (2m_1 f_+ + \Delta E
f_-)v^\mu+(f_+-f_-)r^\mu. $$

A similar analysis applies to the $b\to u$ decays
$B_c\to (D^0,D^{\ast 0})\, e^+\nu$.
Again the heavy quark symmetry analysis is only reliable close
to zero recoil where the u-quark from the $b\to u$ decay has small
momentum. One has

\begin{equation}
\label{b-inv}
M^\mu_b=\frac{\sqrt{2m_P\cdot 2m_H}} {\sqrt{\tilde N_P\cdot
\tilde N_H}} \cdot\frac{1}{m_3}\cdot \int\!\!
\frac{d^4k}{4\pi^2i} {\cal F}_{PH}(-k^2) \frac{{\rm tr}[\gamma^5
\Gamma_H(m_2+\not\! k+\not\! p') O^\mu (1+\not\! v)]}
{[-2kv-2E_1][m_2^2-(k+p')^2]}\,,
\end{equation}
where $q=p-p'=(m_1+E_1-m_H)v-r$. The light quark mass $m_2$ in
Eq.~(\ref{b-inv}) is the u-quark mass.  One finds

\begin{equation}
\label{b-hql}
M^\mu_b=\frac{\sqrt{2m_P\cdot 2m_H}} {\sqrt{\tilde N_P\cdot
\tilde N_H}} \cdot\frac{1}{m_3}\cdot \int\limits_0^\infty\!\!
\frac{dt\, t}{(1+t)^2}\int\limits_0^\infty d\alpha {\cal F}_{PH}(z_b)
\{...\}_{PH}
\end{equation}
with
\begin{eqnarray*}
\{...\}_{PP}&=& \biggl(m_2+\frac{m_H-\alpha t}{1+t}\biggr)\,v^\mu
+\frac{1}{1+t}\,r^\mu\,,
\\
i\{...\}_{PV}&=&
-i\varepsilon^{\mu\nu\alpha\beta}\epsilon^{\ast\nu} v^\alpha r^\beta\,
\biggl(m_2+\frac{m_H-\alpha t}{1+t}\biggr)\,\epsilon^{ast\mu}
-\frac{1}{1+t}\, v^\mu\, \epsilon^{\ast}\cdot r\,.
\end{eqnarray*}
Here,
$
z_b=(\alpha t^2/(1+t))\, (\alpha+2m_H)+t\, (m_2^2-2\alpha E_1)
-(t/(1+t))\,m_H^2.
$
Again, the amplitudes for the $b\to u$ decays are expressed through two
independent functions. The expressions for the form factors in the
heavy quark limit close to zero recoil read

\begin{eqnarray}
F(q^2_{\rm max})&\to& \frac {\sqrt{2m_P\cdot 2 m_H}} {\sqrt{\tilde
N_P\cdot \tilde N_H}} \int\limits_0^\infty\!  \frac{dt \, t}{(1+t)^2}
\int\limits_0^\infty d\alpha
{\cal F}_{PH}(z_c)\{...\}_{F}\,,
\end{eqnarray}
where $F=f_{\pm},A_i,V$ and where

\begin{eqnarray*}
\{...\}_{f_+}&=&-\{...\}_{f_-}=
\frac{1}{2m_3}\cdot\frac{1}{1+t}\,,
\\
\{...\}_{A_1}&=&\frac{1}{m_P+m_V}\frac{1}{m_3}\cdot
\biggl(m_2+\frac{m_H-\alpha t}{1+t}\biggr)\,,
\\
\{...\}_{A_2}&=&-\{...\}_{A_3} =\{...\}_{V}
=\frac{m_P+m_V}{2}\cdot\frac{1}{m_1m_3}\cdot\frac{1}{1+t}\,.
\end{eqnarray*}
Note that one needs to keep the next-to-leading term in the sum
$(f_++f_-)$ to obtain  the above amplitudes.

The hadronic matrix elements of the $b\to c$ decays
$B_c\to (\eta_c, J/\psi) e^+\nu$  simplify significantly in the heavy quark
limit. In this case both b- and c-propagators may be replaced by
their heavy quark limit forms in Eq.~(\ref{b-prop}) with the same
velocity $v$. Again, the results will be valid only near zero recoil.
One has

\begin{equation}
M^\mu_{cc}=\frac{\sqrt{2m_P\cdot 2m_H}} {\sqrt{\tilde N_P \cdot
\tilde N_H}} \cdot\frac{1}{m_3}\cdot \int\!\!
\frac{d^4k}{4\pi^2i} {\cal F}_{PH}(-k^2) \frac{{\rm tr}[\gamma^5
\Gamma_H(1+\not\! v) O^\mu (1+\not\! v)]}
{[-2kv-2E_1][-2kv-2E_2]}\,
\end{equation}
where $p=(m_1+E_1)v$, $p'=(m_3+E_2)v+r$. One finds

\begin{eqnarray}
M^\mu_{cc}&=&\frac{\sqrt{2m_P\cdot 2m_H}} {\sqrt{\tilde N_P \cdot
\tilde N_H}} \cdot\frac{1}{2m_3}\cdot \int\limits_0^\infty\!\!
du\int\limits_0^1 d\alpha {\cal F}_{PH}\biggl(u-2\sqrt{u}\,(\alpha
E_1+(1-\alpha) E_2)\biggr)\{...\}_{PH}
\end{eqnarray}

$$ \{...\}_{PP}= +2v^\mu\,, \hspace{1cm}
i\{...\}_{PV}=-2\epsilon^{\ast\mu}\,. $$ The form factors are written
down

\begin{eqnarray}
F(q^2_{\rm max})&\to& \frac{\sqrt{2m_P\cdot 2m_H}} {\sqrt{\tilde
N_P \cdot \tilde N_H}} \cdot \int\limits_0^\infty\!\!
du\int\limits_0^1 d\alpha {\cal F}_{PH}\biggl(u-2\sqrt{u}\,(\alpha
E_1+(1-\alpha) E_2)\biggr)\{...\}_{F}
\end{eqnarray}
where $F=f_{\pm},A_i,V$. We have

\begin{eqnarray*}
\{...\}_{f_+}&=&\frac{m_1+m_3}{4m_1m_3^2}\,, \hspace{1cm}
\{...\}_{f_-}=-\frac{m_1-m_3}{4m_1m_3^2}\,,
\\
\{...\}_{A_1}&=&\frac{1}{m_P+m_V}\cdot\frac{1}{m_3}\,,
\hspace{0.5cm} \{...\}_{A_2}=-\{...\}_{A_3} =\{...\}_{V}
=\frac{m_P+m_V}{4m_1m_3^2}\,.
\end{eqnarray*}
Thus, our quark loop calculations reproduce the heavy quark limit
relations between form factors obtained in \cite{Jenkins} near zero recoil.
Moreover, we give explicit expressions for the reduced set of form factors
in this limit.

\section{Results and discussion}

Before presenting our numerical results we need to specify our values
for the constituent quark masses and shapes of the vertex functions.
As concerns the vertex functions, we found a good description of various
physical quantities \cite{IS} adopting a Gaussian form for them.
Here we apply the same procedure using $\phi_H(k^2)=\exp\{-k^2/\Lambda_H^2\}$
in the Euclidean region. The magnitude of $\Lambda_H$ characterizes  the size
of the vertex function and is an adjustable parameter in our model.
We reiterate that all the analytical results presented in Sec.V are valid
for any choice of form factor $\phi_H(k^2)$. For example,
we have reproduced  the results of \cite{NW} where dipole form factor
was adopted by using our general formula.

In \cite{IS} we have studied various decay modes of the $\pi$, $K$, $D$, $D_s$,
$B$ and $B_s$ mesons. The  $\Lambda$-parameters and the
constituent quark masses were determined by a least-squares fit to experimental
data and lattice determinations.
The obtained values for the charm and bottom quarks (see Eq.~(\ref{fitmas})
allow us to consider the low-lying charmonium ($\eta_c$ and $J/\psi$) and
bottonium ($\Upsilon$) states, and also the new-observed $B_c$-meson.

\begin{equation}
\begin{array}{ccccccc}
m_u & &  m_s & &  m_c & & m_b \\
\hline
$ 0.235 $ & & $ 0.333 $
& &  $ 1.67 $ & & $ 5.06 $  \\
\end{array}
\label{fitmas}
\end{equation}

Basically we use either the available experimental values or
the values of lattice simulations for the leptonic decay constants
to adjust the size parameters $\Lambda_H$.
The value of $f_{B_c}$ is unknown and theoretical predictions
for it lie within the 300-600 MeV range. We choose
the value of $f_{B_c}=360$ MeV, being the average QCD sum rule
predictions, for fitting $\Lambda_{B_c}$.
The obtained  values
of $\Lambda_H$ are listed in Eq.~(\ref{fitlam}) as well as the
values of $f_H$ in Table I.

\begin{equation}\label{fitlam}
\begin{array}{ccccccccc}
\Lambda_\pi & \Lambda_K & \Lambda_D & \Lambda_{D_s} &
\Lambda_{J/\psi} & \Lambda_B & \Lambda_{B_s} & \Lambda_{B_c}&
\Lambda_{\Upsilon }\\ \hline $\ \ 1.16\ \ $ & $\ \ 1.82\ \ $ & $\
\ 1.87\ \ $ & $\ \ 1.95\ \ $ & $\ \ 2.12\ \ $ & $\ \ 2.16\ \ $ &
$\ \ 2.27\ \ $ &$\ \ 2.43\ \ $  & $\ \ 4.425\ \ $  \\
\end{array}
\end{equation}

The values of $\Lambda_H$ are such that
$\Lambda_{m_i}<\Lambda_{m_j}$ if $m_i<m_j$. This corresponds to
the ordering law for sizes of bound heavy-light states.

The situation with the determination of $\Lambda_{\eta_c}$ is
quite unusual. Naively one expects that $\Lambda_{\eta_c}$ should be
the same as $\Lambda_{J/\psi}$. However, in this case the value of the
$\eta_c\to\gamma\gamma$ decay width comes out to be 2.5 less than
the experimental average. The experimental average can be reached
only for relatively large value of $\Lambda_{\eta_c}=4.51$ GeV.
Note that the values of the other observables ($J/\psi\to\eta_c\gamma$
 and $B_c\to\eta_c l \nu$ decay rates) are not so sensitive to the choice
of $\Lambda_{\eta_c}$:

\begin{eqnarray*}
{\rm Br}(\eta_c\to \gamma\ \gamma)\,\,& =& 0.031 \,(0.012)\,\% \,,
\hspace{0.7cm} {\rm expt.}=(0.031 \pm 0.012)\,\%\,,
\\
{\rm Br}(J/\psi\to\eta_c\gamma) & =& 0.90\, (1.00)\ \%
\,,\hspace{1cm} {\rm expt.} = (1.3 \pm 0.4)\, \% \,,
\\
{\rm Br}(B_c\to \eta_c l\nu) & =& 0.98\, (1.02)\ \% \,.
\end{eqnarray*}
The values in parenthesis correspond to the case of equal sizes
for the charmonium states.

We  concentrate  our study on the semileptonic decays of
the $B_c$-meson. To extend the number of modes, we consider also
the decays into the vector mesons $D^\ast$, $B^\ast$ and $B_s^\ast$.
We will use the masses and sizes of their pseudoscalar partners
for the numerical evaluation of the form factors  to avoid the
appearance of imaginary parts in the amplitudes. Such an assumption
is justified by the small differences of their physical masses.

\begin{figure}[t]
\epsfxsize=10cm \centerline{\epsffile{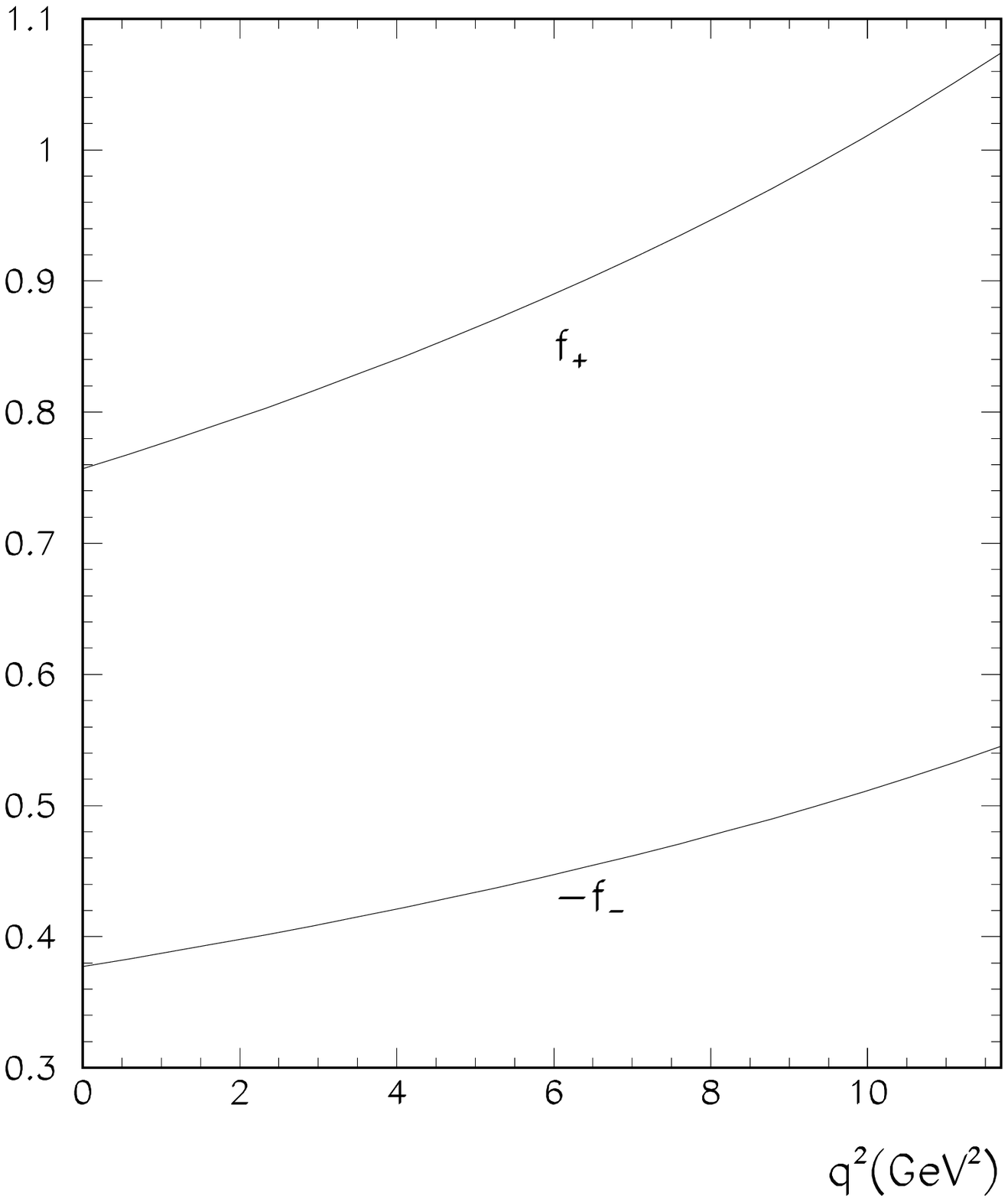}}
\caption{$q^2$-dependence of the $B_c\rightarrow \eta_c$
form factors. Note that we plot the negative of the $f_-(q^2)$ form factor.}
\label{f:bcetac}
\end{figure}

\begin{figure}[t]
\epsfxsize=10cm \centerline{\epsffile{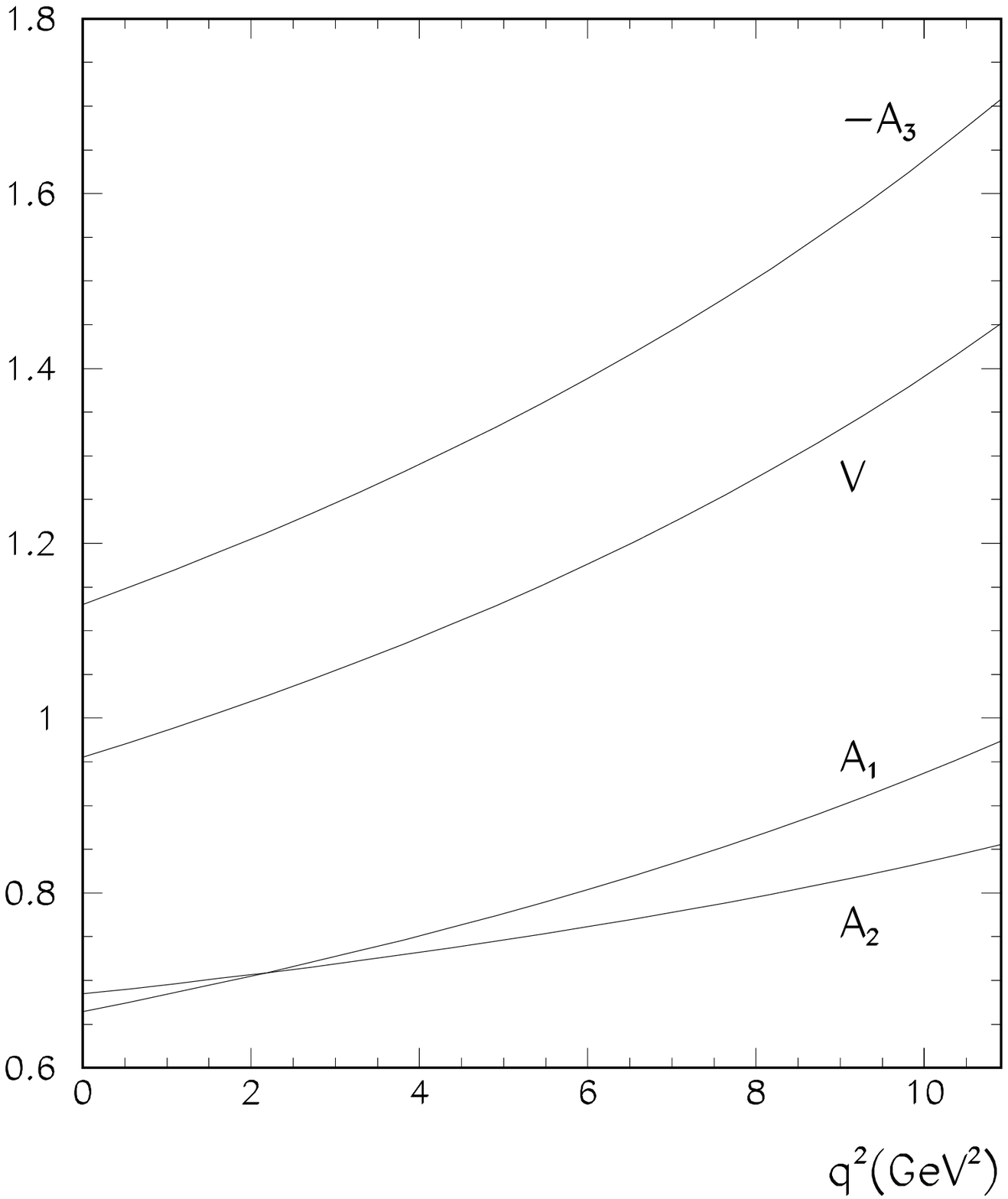}}
\caption{$q^2$-dependence of the $B_c\rightarrow J/\psi $
form factors. Note that we plot the negative of the $A_3(q^2)$ form factor. }
\label{f:bcjp}
\end{figure}

\begin{figure}[t]
\epsfxsize=10cm \centerline{\epsffile{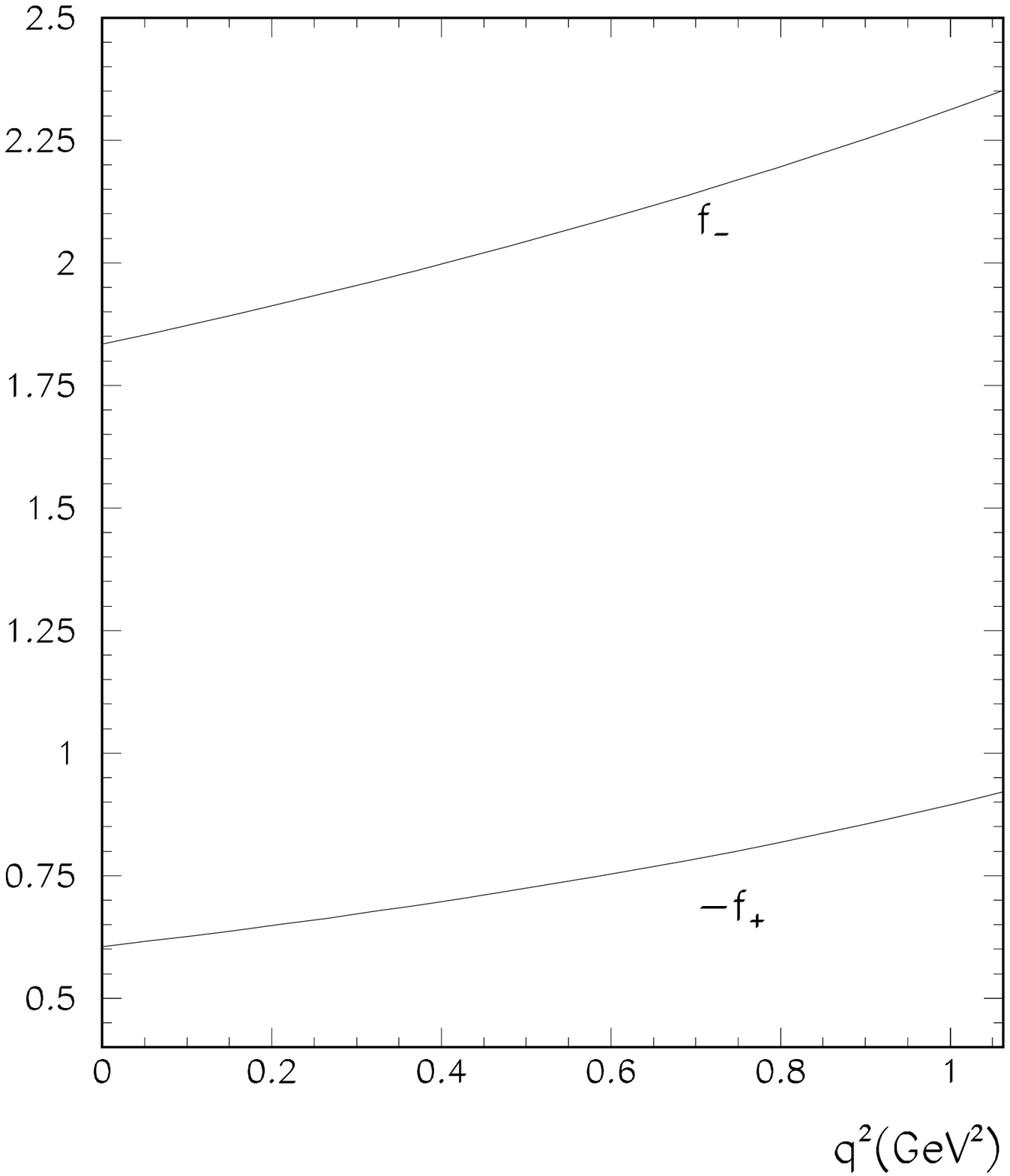}}
\caption{$q^2$-dependence of the $B_c\rightarrow B_s$
form factors. Note that we plot the negative of the $f_+(q^2)$ form factor.}
\label{f:bcbs}
\end{figure}

\begin{figure}[ht]
\epsfxsize=10cm \centerline{\epsffile{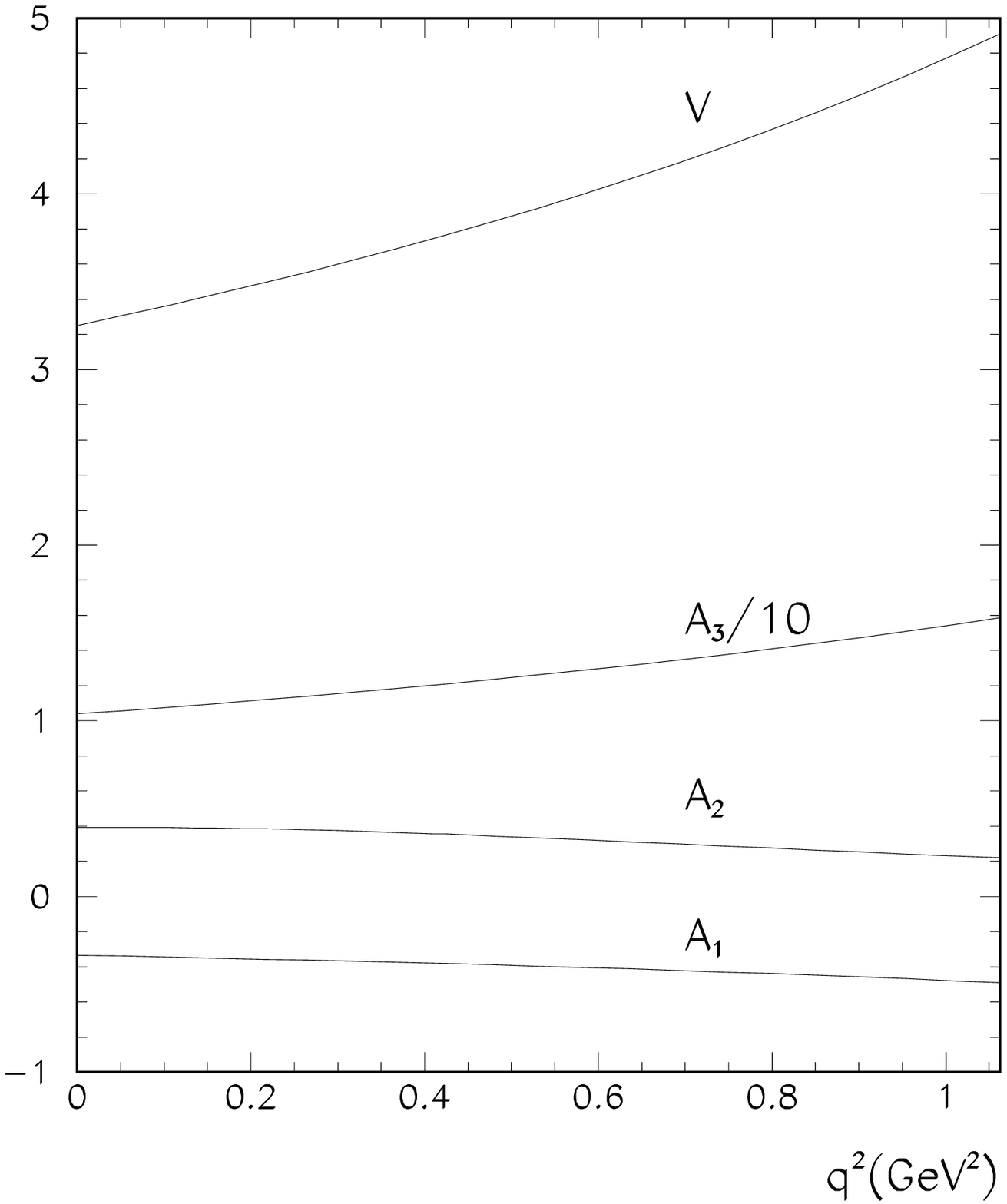}}
\caption{$q^2$-dependence of the $B_c\rightarrow B^*_s $
form factors.  }
\label{f:bcbsstar}
\end{figure}

In Figs.~(\ref{f:bcetac}-\ref{f:bcbsstar}) we show the calculated
$q^2$ dependence in the full physical regions of the semileptonic
form factors of the CKM-enhanced transitions
$B_c\to \eta_c$, $B_c\to J/\psi$ and $B_c\to B_s$, $B_c\to B_s^\ast$.
The values of form factors at maximum and zero recoil are listed
in Tables II-IV. The comparison of the exact values of form
factors at zero recoil and those obtained in the heavy quark limit
is given in Table V.
 Our results indicate that the corrections to the heavy quark limit
 at the zero recoil point $q^2=q^2_{\rm max}$ can be as large as
 a factor of two in $b-c$ transitions and a factor of almost five
 in $b-u$ and $c-d$ transitions. This is not so surprising considering the
 semileptonic decays of the $D$ meson where similar corrections can amount
 to a factor of two \cite{DSEH}.

The form factors can be {\it approximated} by the form

\begin{equation}
\label{appr_form}
f(q^2)=\frac{f(0)}{1-q^2/m^2_{\rm fit}-\delta\cdot (q^2/m^2_{\rm fit})^2}
\end{equation}
with the dimensionless values of $f(0)$ given in Tables II and III.
Note that the form factor $f_+(q^2)$ for the $B_c\to\eta_c$ transition
rises with $q^2$ as is appropriate in the time-like region. When plotted
against $\omega=p\cdot p'/(m_{\rm in}\cdot m_{\rm out})=
(m_{\rm in}^2+m_{\rm out}^2-q^2)/(2\,m_{\rm in}\cdot m_{\rm out})$,
they would fall with $\omega$ as one is familiar with heavy quark effective
theory.

It is interesting that the obtained values  of $m^2_{\rm fit}$ for
the CKM-enhanced transitions (see Table~VIII)
are very close to the values of the appropriate lower-lying
($\bar q q'$) vector mesons
($m_{B_c^\ast}\approx m_{B_c} =6.4$ GeV for $b\to c$,
 $m_{D_s^\ast}=2.11$ GeV for $c\to s$.).
The parameter $\delta$ characterizes the admixture of a $q^4$-term in the
denominator. Its magnitude is relatively small for all form factors of
the CKM-enhanced transitions except  $A_1$ for $B_c\to J/\psi$
transition and $A_2$ for $B_c\to B_s^\ast$ which have a rather flat
behavior. This means that those form factors  can be
reliably approximated by a vector dominance form. However,
one cannot approximate the form factors for the CKM-suppressed transitions
by a  pole-like function only.

We use the calculated form factors in Eq.~(\ref{width}) to evaluate the
branching ratios for various semileptonic $B_c$ decay modes including
their $\tau$-modes when they are kinematically accessible.
We report the calculated values of a wide range of branching ratios in
Table VI. The results of other  approaches are also given for comparison.
The values of branching ratios of the CKM-enhanced modes
with an electron in the final state are of order 1-2 $\%$.
The values of branching ratios  of the CKM-suppressed modes are considerably
less. The modes with a $\tau$-lepton in the final state are  suppressed
due to the reduced phase space in these modes.
To complete our predictions for the physical observables
we give in Table VII the values of the polarization ratio and
forward-backward asymmetry for the prominent decay modes.

\section*{Acknowledgments}

We would like to thank  F. Buccella for many interesting discussions.
M.A.I. gratefully acknowledges the hospitality of the Theory
Groups at Mainz and Naples Universities where  this work was
completed. His visit at Mainz University was supported  by the DFG
(Germany) and the Heisenberg-Landau fund. J.G.K. acknowledges
partial support by the BMBF (Germany) under contract 06MZ865.

\section{Appendix}
In this Appendix we list the remaining form factor expressions appearing
in the curly brackets in Eq.~(\ref{ff}) which have not been listed in
the main text.

\noindent {\bf b-decay:}

\begin{eqnarray*}
\biggl\{...\biggr\}^b_{f_-}&=&
-{\cal F}_{PP}(z_b) \frac{1}{1+t}
\biggl[3(\alpha_1-\alpha_2)\frac{t}{1+t}\biggr]
 +{\cal F}'_{PP}(z_b) \biggl\{ (m_1-m_2)\,m_3
\\
&& +\frac{t}{1+t} \biggl(
(\alpha_1-\alpha_2)(m_1m_3+m_2m_3-m_1m_2) +\alpha_1 p^2-\alpha_2
p'^2 \biggr) \\ && -(\alpha_1-\alpha_2)\biggl[(\alpha_1+\alpha_2)
(\alpha_1\, p^2 +\alpha_2 \,p'^2) -\alpha_1\alpha_2\,
q^2\biggr]\left(\frac{t}{1+t}\right)^3 \biggr\}
 \\
&& \\ \biggl\{...\biggr\}^b_{A_1}&=& \frac{2}{m_P+m_V} \biggl\{
 {\cal F}(z_b)_{PV}\,\frac{1}{1+t} (m_1+2m_2-m_3)
 -{\cal F}'_{PV}(z_b)
\biggl[ m_1m_2m_3+\frac{1}{2}(p^2+p'^2-q^2)\,m_3
\\
&& +\frac{1}{2}\frac{t}{1+t} \biggl( \left(\alpha_1
m_1+(2\alpha_1+\alpha_2)\,m_2
             -(3\alpha_1+\alpha_2)m_3\right)\,p^2
\\
&& +\left((\alpha_1+2 \alpha_2)m_1+\alpha_2 m_2
             -(\alpha_1+3\alpha_2)\,m_3\right)\,p'^2
\\
&& -\left(\alpha_1 m_1+\alpha_2
m_2-(\alpha_1+\alpha_2)\,m_3\right)\,q^2 \biggr)
\\
&& -(m_1+m_2-m_3) \biggl( (\alpha_1+\alpha_2)(\alpha_1
\,p^2+\alpha_2 \,p'^2) -\alpha_1\alpha_2 \,q^2 \biggr)
\left(\frac{t}{1+t}\right)^2 \biggr] \biggr\}.
\\
&&\\
\biggl\{...\biggr\}^b_{A_2}&=&
(m_P+m_V) \biggl[ {\cal F}'_{PV}(z_b)\biggr]
\biggl\{ -m_3-\frac{t}{1+t} \biggl[ \alpha_1
m_1+\alpha_2 m_2 -(3\alpha_1+\alpha_2)\,m_3 \biggr]
\\
&& +2(m_1-m_3)\,\alpha_1\,(\alpha_1+\alpha_2)
\left(\frac{t}{1+t}\right)^2 \biggr\}.
\\
&&\\
\biggl\{...\biggr\}^b_{A_3}&=& (m_P+m_V)
\biggl[ {\cal F}'_{PV}(z_b)\biggr]
\biggl\{ m_3+\frac{t}{1+t} \biggl[ \alpha_1
m_1+\alpha_2 m_2 +(\alpha_1-\alpha_2)\,m_3 \biggr]
\\
&& +2(m_1-m_3)\,\alpha_1\,(\alpha_1-\alpha_2)
\left(\frac{t}{1+t}\right)^2 \biggr\}.
\\
&&\\
\biggl\{...\biggr\}^b_V&=&
(m_P+m_V) \biggl[ -{\cal F}'_{PV}(z_b)\biggr]
\biggl\{ m_3+\frac{t}{1+t} \biggl[ \alpha_1
(m_1-m_3)+\alpha_2 (m_2 -m_3) \biggr] \biggr\}.
\end{eqnarray*}
We use the abbreviations
$$ P_b=\alpha_1 p+\alpha_2 p'\,, \hspace{1cm}
z_b=t\,\left(\sum\limits_{i=1}^3 \alpha_i m^2_i-\alpha_1\alpha_3 p^2
-\alpha_2\alpha_3 p'^2-\alpha_1\alpha_2 q^2\right)-\frac{t}{1+t}\, P^2_b .
$$

\noindent {\bf c-decay:}

\begin{eqnarray*}
\biggl\{...\biggr\}^c_{f_-}&=&
3\,{\cal F}_{PP}(z_c) \frac{1}{1+t}
\biggl[1-(\alpha_1+2\alpha_2)\frac{t}{1+t}\biggr]
 +{\cal F}'_{PP}(z_c)
\biggl\{ -2 m_1 m_3 + m_2 m_3
\\
&& +\frac{t}{1+t} \biggl(
(\alpha_1+2\alpha_2)(m_1m_2+m_1m_3-m_2m_3) -2(\alpha_1+\alpha_2)\,
p^2 +2\alpha_2\, p'^2-\alpha_2\, q^2 \biggr)
\\
&& +\left(\frac{t}{1+t}\right)^2
\biggl((3\alpha_1^2+6\alpha_1\alpha_2+2\alpha_2^2)\, p^2
-2\alpha_2(\alpha_1+\alpha_2)\,
p'^2+\alpha_2(2\alpha_1+3\alpha_2)\, q^2\biggr)
\\
&& -\left(\frac{t}{1+t}\right)^3 (\alpha_1+2\alpha_2) \biggl(
\alpha_1(\alpha_1+\alpha_2)\, p^2 -\alpha_1\alpha_2\,
p'^2+\alpha_2 (\alpha_1+\alpha_2)\,q^2\biggr).
\\
&& \\
\biggl\{...\biggr\}^c_{A_1}&=&
\frac{2}{m_P+m_V}
\biggl\{ {\cal F}_{PV}(z_c)\frac{1}{1+t} (m_1-2m_2-m_3)
 -{\cal F}'_{PV}(z_c)
\biggl[ -m_1m_2m_3+\frac{1}{2}(p^2+q^2-p'^2)\,m_3
\\
&& +\frac{1}{2}\frac{t}{1+t} \biggl( \left( \alpha_1
m_1-(2\alpha_1+\alpha_2)\,m_2
             -(3\alpha_1+\alpha_2)m_3\right)\,p^2
\\
&& +\left(-\alpha_1 m_1+\alpha_2 m_2
             +(\alpha_1+\alpha_2)\,m_3\right)\,p'^2
\\
&& +\left( (\alpha_1+2\alpha_2) m_1-\alpha_2
m_2-(\alpha_1+3\alpha_2)\,m_3 \right)\,q^2 \biggr)
\\
&& -\left(\frac{t}{1+t}\right)^2\,(m_1-m_2-m_3)
\biggl(\alpha_1(\alpha_1+\alpha_2)\,p^2+\alpha_2(\alpha_1+\alpha_2)\,q^2
-\alpha_1\alpha_2\,p'^2\biggr) \biggr\}.
\\
&&\\
\biggl\{...\biggr\}^c_{A_2}&=&
(m_P+m_V)
\biggl[ -{\cal F}'_{PV}(z_c)\biggr]
\biggl\{ m_3+\frac{t}{1+t}
\biggl[ \alpha_1 m_1-\alpha_2 m_2 -(3\alpha_1+\alpha_2)\,m_3 \biggr]
\\
&& -2(m_1-m_3)\,\alpha_1\,(\alpha_1+\alpha_2)
\left(\frac{t}{1+t}\right)^2 \biggr\}.
\\
&&\\
\biggl\{...\biggr\}^c_{A_3}&=&
(m_P+m_V)
\biggl[ {\cal F}'_{PV}(z_c)\biggr]
\biggl\{ -3m_3+\frac{t}{1+t}
\biggl[-(3\alpha_1+4\alpha_2) m_1-\alpha_2 m_2
+(5\alpha_1+7\alpha_2)\,m_3 \biggr]
\\
&& +2(m_1-m_3)\,(\alpha_1+2\alpha_2)\,(\alpha_1+\alpha_2)
\left(\frac{t}{1+t}\right)^2 \biggr\}.
\\
&&\\
\biggl\{...\biggr\}^c_V&=&
(m_P+m_V)
\biggl[ -{\cal F}'_{PV}(z_c)\biggr]
\biggl\{ m_3+\frac{t}{1+t}
\biggl[ \alpha_1 (m_1-m_3)+\alpha_2 (m_2 -m_3) \biggr] \biggr\}.
\end{eqnarray*}
Here we have used the abbreviations
$$ P_c=\alpha_1 p+\alpha_2 q\,, \hspace{1cm}
z_c=t\,\left(\sum\limits_{i=1}^3 \alpha_i m^2_i-\alpha_1\alpha_3\, p^2
-\alpha_2\alpha_3\, q^2-\alpha_1\alpha_2\, p'^2\right)
-\frac{t}{1+t}\, P_c^2 .$$

\newpage

\begin{table}[t1]
\caption {Leptonic decay constants $f_H$ (MeV) used in the least-squares fit.}
\begin{center}
\begin{tabular}{|l|l|ll|}
\hline Meson & This model & Other  & Ref. \\
\hline\hline
$\pi^+$& 131& $130.7\pm 0.1\pm 0.36$       & Expt. \cite{PDG} \\
\hline
 $K^+$ & 160 & $159.8\pm 1.4\pm 0.44$       &
Expt.  \cite{PDG} \\
\hline
   $D^+$ & 191 &
$191^{+19}_{-28}$   & Lattice \cite{FS,Wit} \\
          &  & $192\pm 11^{+16+15}_{-8-0} $ & Lattice \cite{MILC} \\
          &  & $194^{+14}_{-10}\pm 10$      & Lattice \cite{Fermi} \\
\hline $\,\, D^+_s$ & 206    & $206^{+18}_{-28}$         & Lattice
\cite{FS,Wit} \\
        &        & $210\pm 9^{+25+17}_{-9-1}$   & Lattice \cite{MILC} \\
        &        & $213^{+14}_{-11}\pm 11$      & Lattice \cite{Fermi} \\
\hline   $B^+$& 172 & $172^{+27}_{-31}$    & Lattice
\cite{FS,Wit} \\
          &  & $157\pm 11^{+25+23}_{-9-0} $ & Lattice \cite{MILC} \\
          &  & $164^{+14}_{-11}\pm 8 $      & Lattice \cite{Fermi} \\
\hline
  $\,\, B^+_s$& 196   & $171\pm 10^{+34+27}_{-9-2} $ & Lattice \cite{MILC} \\
         &   & $185^{+13}_{-8}\pm 9 $      & Lattice \cite{Fermi} \\
\hline
 $\,\, B_c$  &  & 479         & logarithmic potential \cite{EQ} \\
        &  & 500         & Buchm\"uller-Tye potential \cite{EQ} \\
        &  & 512         & power-law potential \cite{EQ} \\
        &  & 687         & Cornell potential \cite{EQ} \\
        &  & 480         & Potential model \cite{CC} \\
        &  & 432         & QCD-inspired QM \cite{CF2} \\
        &  & 400$\pm$ 20 & QCD spectral SR \cite{Nar} \\
        &  & 300         & QCD SR \cite{AY} \\
        &  & 360$\pm$ 60 & QCD SR \cite{CNP} \\
        &  & 300$\pm$ 65 & QCD SR \cite{Ch} \\
        &  & 385$\pm$ 25 & QCD SR \cite{Kis} \\
        &  & 420(13)     & Lattice NRQCD  \cite{JW} \\
\hline\hline   $B_c$ &360  & 360         & Our average
of QCD SR\\
%
%
\hline\hline  $J/\psi$ &404   & 405$\pm$ 17   & Expt.
\cite{PDG} \\ \hline  $\Upsilon$ & 711 & 710$\pm$37 &
Expt. \cite{PDG}\\ \hline
\end{tabular}
\end{center}
\end{table}

\begin{table}[t2]
\caption {Predictions for the form factors at $q^2=0$ and
$q^2=q^2_{\rm max}$ for $B_c\to P$ decays. }
\begin{center}
\begin{tabular}{|c|c|c|c|c|}
\hline
 P & $f_+(0)$ & $f_+(q^2_{\rm max})$ & $f_-(0)$ & $f_-(q^2_{\rm max}) $ \\
\hline\hline
 $\eta_c$ &  0.76 & 1.07  & -0.38 & -0.55 \\
\hline
  D       & 0.69 & 2.20 &-0.64 & -2.14 \\
\hline\hline
  B       &-0.58 & -0.96 & 2.14 & 2.98 \\
\hline
 $B_s$    &-0.61 &-0.92& 1.83& 2.35 \\
\hline
\end{tabular}
\end{center}
\end{table}

\begin{table}[t1]
\caption {Predictions for the form factors at $q^2=0$ for $B_c\to
V$ decays. }
\begin{center}
\begin{tabular}{|c|c|c|c|c|}
\hline
 V          & $A_1(0)$ & $A_2(0)$ & $A_3(0)$ & $V(0)$ \\
\hline\hline
 $J/\psi$   &  0.68    &  0.66    & -1.13     &0.96 \\
\hline
 $D^\ast$   &  0.56    &  0.64   & -1.17    &0.98 \\
\hline\hline
 $B^\ast$   & -0.27    & 0.60   &  10.8    & 3.27 \\
\hline
 $B_s^\ast$ & -0.33    & 0.40   & 10.4    &3.25\\
\hline
\end{tabular}
\end{center}
\end{table}

\begin{table}[t1]
\caption {Predictions for the form factors at $q^2=q^2_{\rm max}$
for $B_c\to V$ decays. }
\begin{center}
\begin{tabular}{|c|c|c|c|c|}
\hline
 V         & $A_1(q^2_{\rm max})$ & $A_2(q^2_{\rm max})$ &
             $A_3(q^2_{\rm max})$ & $V(q^2_{\rm max})$ \\
\hline\hline
 $J/\psi$   &  0.86 &  0.97    & -1.71     & 1.45 \\
\hline
 $D^\ast$   &  0.85 & 1.76   &-3.69    & 3.26 \\
\hline\hline
 $B^\ast$   &-0.42&  0.49  &  18.0    & 5.32 \\
\hline
 $B_s^\ast$ & -0.49 & 0.21   & 15.9  &4.91 \\
\end{tabular}
\end{center}
\end{table}

\begin{table}[t1]
\caption {Comparison of the form factors at the zero recoil point
$q^2=q^2_{\rm max}$
calculated in the heavy quark limit with exact results. }
\begin{center}
\begin{tabular}{|c|c|c|c|c|c|c|}
\hline
 H                       & $f_+$ &$f_-$  & $A_1$ & $A_2$  & $A_3$ & $V$  \\
\hline\hline
 $\eta_c$,$J/\psi$       &  1.07 & -0.55 & 0.86  &  0.97  &-1.71  & 1.45 \\
 $\eta_c$,$J/\psi$ (HQL) &  0.70 & -0.35 & 0.37  &  0.69  &-0.69  & 0.69 \\
\hline
 D, $D^\ast$             &  2.20 & -2.14 & 0.85  & 1.76   &-3.69  & 3.26 \\
 D, $D^\ast$ (HQL)       & 0.59  & -0.59 & 0.18  & 1.50   &-1.50  & 1.50 \\
\hline
 B, $B^\ast$             & -0.96 & 2.98  &-0.42  & 0.49   & 18.0  & 5.32 \\
 B, $B^\ast$ (HQL)       & -0.47 & 1.67  &-0.25  & 1.91   & 21.47 & 1.91 \\
\hline
\end{tabular}
\end{center}
\end{table}

\begin{table}[t]
\caption{ Branching ratios BR($\%$) for the  semileptonic decays
$B_c^+ \rightarrow H l^+ \protect\nu $, calculated with
the CDF central value $\protect\tau (B_c)=0.46$ ps $\protect\cite{CDF}$.}
\begin{center}
\begin{tabular}{|c|c|c|c|c|c|c|c|}
\hline\hline H & This model & \cite{KLO,KKL}
&\cite{AMV}&\cite{CC}&\cite{CF1}&\cite{AKNT} &\cite{NW}\\
\hline\hline
$\eta_c\, e\,\nu$ &0.98 & 0.8$\pm$0.1 & 0.78& 1.0 & 0.15(0.5) & 0.6 & 0.52 \\
\hline
$\eta_c\,\tau\,\nu$   & 0.27   & & & &  &  &  \\
\hline
$ J/\psi\, e\,\nu$   & 2.30  &
2.1$\pm$0.4 & 2.11 & 2.4 & 1.5(3.3) & 1.2 & 1.47\\
\hline
$J/\psi \,\tau\,\nu$    & 0.59   & &  & &  &  &  \\
\hline
$ D^0\, e\,\nu$  & 0.018  &     & 0.003 & 0.006 & 0.0003(0.002) & &  \\
\hline
$ D^0\,\tau\,\nu$    & 0.0094  &  & &  &  & & \\
\hline
$ D^{\ast 0} \, e\,\nu$ & 0.034  &      & 0.013 & 0.019 &0.008(0.03) & &  \\
$D^{\ast 0} \,\tau\,\nu$    & 0.019 &  & &  &  & & \\
\hline\hline
$B^0\, e\,\nu $     &0.15  &       & 0.08 & 0.16 & 0.06(0.07) & & \\
\hline
$ B^{\ast 0}\, e\,\nu$   & 0.16  &     &0.25 & 0.23 &0.19(0.22)  & &    \\
\hline
$ B^0_s\, e\,\nu $    & 2.00  & 4.0 & 1.0 & 1.86 & 0.8(0.9) & 1.0 & 0.94 \\
\hline $ B_s^{\ast 0}\, e\,\nu$ & 2.6   & 5.0  & 3.52 & 3.07  &2.3(2.5)  & & 1.44 \\
\hline
\end{tabular}
\end{center}
\end{table}

\begin{table}[t]
\caption{ The polarization ratio $\alpha$ and forward-backward
asymmetry $A_{FB}$.}
\begin{center}
\begin{tabular}{|c|c|c|}
\hline\hline H               & $\alpha$ & $A_{FB}$ \\
\hline\hline
$ J/\psi$       & 1.15     & -0.21  \\
\hline
$ D^{\ast 0}$   & 0.10     & -0.46   \\
\hline
$ B^{\ast 0}$   & 0.94     & 0.35 \\
\hline
$ B_s^{\ast 0}$ & 1.09     & 0.29  \\
\hline
\end{tabular}
\end{center}
\end{table}

\begin{table}[t]
\caption{The numerical values of $m^2_{\rm fit}$
(GeV$^2$) and $\delta$ in the form factor parameterization
Eq.~(\ref{appr_form}).}
\begin{center}
\begin{tabular}{|ll|cccccc}
\hline
                      && $f_+$ & $f_-$ & $A_1$ & $A_2$ & $A_3$ & V \\
\hline
$B_c\to\eta_c,J/\psi$ &
$m^2_{\rm fit}$&$(6.37)^2$&$(6.22)^2$&$(8.20)^2$&$(5.91)^2$&$(5.67)^2$&
$(5.65)^2 $  \\
 & $\delta$         & 0.087 &0.060 & 1.40 & 0.052 &-0.004 &0.0013 \\
\hline
$B_c\to B_s,B_s^\ast$  &
$m^2_{\rm fit}$   & $(1.73)^2$  & $(2.21)^2$ &$(1.86)^2$&$(3.44)^2$&
$(1.73)^2$&$(1.76)^2$ \\
 & $\delta$      &-0.09 &0.07 &0.13&-107&-0.09&-0.052  \\
\hline
\end{tabular}
\end{center}
\end{table}

\newpage

\newpage

\newpage

\newpage

\newpage

\begin{references}
%

\bibitem{CDF}  CDF Collaboration, F. Abe {\it et al.} Phys. Rev. {\bf D58},
112004 (1998); Phys. Rev. Lett. {\bf 81}, 2432 (1998).
%
%
\bibitem{IW}  N. Isgur, M.B. Wise, Phys. Lett. {\bf B232}, 113 (1989); Phys.
Lett. {\bf B237}, 527 (1990).
%
%
\bibitem{static}  B.A. Thacker, G.P. Lepage, Phys. Rev. {\bf D43}, 196
(1991).
\bibitem{Jenkins}  E. Jenkins, M. Luke, A.V. Manohar, M.J. Savage, Nucl.
Phys. {\bf B390}, 463 (1993). %
%
%

\bibitem{Gersh}  S.S. Gershtein {\it et al.}, "Theoretical status of the
$B_c $ meson", hep-ph/9803433; in Proc. ``Progress in Heavy Quark
Physics'', eds.: M. Beyer, T. Mannel, H. Schroder, (University of
Rostock, Germany), 1998, p. 272.
%
%
\bibitem{CC} C.-H. Chang and Y.-Q. Chen, Phys. Rev. {\bf D49}, 3399 (1994).
\bibitem{AMV}  A. Abd El-Hady, J. H. Mu$\tilde {{\rm n}}$oz and J.P. Vary,
Phys. Rev. {\bf D62}, 014019 (2000).
\bibitem{AKNT}  A.Yu. Anisimov, P.Yu. Kulikov, I.M. Narodetskii,
K.A. Ter-Martirosyan,\newline Phys. Atom. Nucl. {\bf 62}, 1739
(1999).
%
\bibitem{KLO}  V.V. Kiselev, A.K. Likhoded, A.I. Onishchenko,
Nucl. Phys. {\bf B259}, 473 (2000).
\bibitem{KKL}  V.V. Kiselev, A.E. Kovalsky, A.K. Likhoded,
``$B_c$-decays and lifetime in QCD Sum Rules'', hep/ph-0002127.
\bibitem{CF1} P. Colangelo and F. De Fazio, Phys. Rev. {\bf D61}, 034012, 2000.
%
%
\bibitem{RCQM}  M.A. Ivanov, M.P. Locher and V.E. Lyubovitskij, Few-Body
Syst. {\bf 21}, 131 (1996); \newline M.A. Ivanov and V.E.
Lyubovitskij, Phys. Lett. {\bf B408}, 435 (1997).
%
%
\bibitem{SWH}  A. Salam, Nuovo Cim. {\bf 25}, 224 (1962); S. Weinberg, Phys.
Rev. {\bf 130}, 776 (1963);\newline K. Hayashi {\it et al.}, Fort.
der Phys. {\bf 15}, 625 (1967).
%
%
\bibitem{EI}  G.V. Efimov and M.A. Ivanov, Int. J. Mod. Phys. {\bf A4}, 2031
(1989); \newline ``{\it The Quark Confinement Model of Hadrons}",
IOP Publishing, 1993.
%
%
\bibitem{DSE}  C.D. Roberts and A.G. Williams,
Prog. Part. Nucl. Phys. {\bf 33}, 477 (1994).
\bibitem{DSEH}  M.A. Ivanov, Yu.L. Kalinovsky, C.D. Roberts,
Phys. Rev. {\bf D60}, 034018 (1999);
 M.A. Ivanov, Yu.L. Kalinovsky, P. Maris, C.D. Roberts,
Phys. Lett. {\bf B416}, 29 (1998); Phys. Rev. {\bf C57}, 1991
(1998).
%
%
\bibitem{IS}  M.A. Ivanov and P. Santorelli, Phys. Lett. {\bf B456}, 248
(1999).
%
%
\bibitem{RTQM}  M.A. Ivanov, V.E. Lyubovitskij, J.G. K\"{o}rner, P. Kroll,
Phys. Rev. {\bf D56}, 348 (1997); M.A. Ivanov, J.G. K\"{o}rner,
V.E. Lyubovitskij, A.G. Rusetsky, Phys. Rev. {\bf D57}, 5632 (1998);
Phys. Rev. {\bf D60}, 094002 (1999); Phys. Rev. {\bf D61}, 114010
(2000).
\bibitem{RTQM1}  M.A. Ivanov, J.G. K\"{o}rner, V.E. Lyubovitskij,
A.G. Rusetsky,\newline Phys. Lett. {\bf B476}, 58 (2000).
%
%
\bibitem{Gatto}  A. Deandrea, N. Di Bartolomeo, R. Gatto, G. Nardulli and
A.D. Polosa, \newline Phys. Rev. {\bf D58}, 034004 (1998).
%
%
\bibitem{NW} M.A. Nobes and R.M. Woloshyn, J. Phys. {\bf G26}, 1079 (2000).
%
%
\bibitem{KS} J.G. K\"orner and G.A. Schuler, Z. Phys. {\bf C46}, 93 (1990).

\bibitem{FS} J.M. Flynn, C.T. Sachrajda, ``Heavy Quark Physics
from Lattice QCD'', hep-lat/9710057; In Proc. ``Heavy Flavors
II'', eds.: A.J. Buras, M. Linder, (WS, Singapore), 1997, 402-452.
\bibitem{Wit} H. Wittig, Int. J. Mod. Phys. {\bf A12} 4477 (1997).
\bibitem{MILC} C. Bernard {\it et al.}, Phys. Rev. Lett. {\bf 81},
               4812 (1998).
\bibitem{Fermi} A.X. El-Khadra {\it et al.}, Phys. Rev. {\bf D58},
014506 (1998).
%
%
\bibitem{EQ} E.J. Eichten and C. Quigg, Phys. Rev. {\bf D49}, 5845 (1994).
\bibitem{CF2} P. Colangelo and F. De Fazio, Mod. Phys. Lett. {\bf A14},
2303, (1999).
\bibitem{Nar} S. Narison, Phys. Lett. {\bf B210}, 238 (1988).
\bibitem{AY} T.M. Aliev and O. Yilmaz, Nuovo Cim. {\bf A105}, 827 (1992).
\bibitem{CNP} P. Colangelo, G. Nardulli and N. Paver,
Z. Phys. {\bf C57}, 43 (1993).
\bibitem{Ch} M. Chabab, Phys. Lett. {\bf B325}, 205 (1994).
\bibitem{Kis} V.V. Kiselev, Int. J. Mod. Phys. {\bf A11}, 3689 (1996).
\bibitem{JW} B.D. Jones and R.M. Woloshin, Phys. Rev. {\bf D60}, 014502
(1999).
\bibitem{PDG} C. Caso {\it et al.} (Particle Data Group), European
Phys. Journal {\bf C3}, 1 (1998).
\end{references}
\end{document}